\documentclass[sigconf,screen, review=false,timestamp=false]{acmart}
\AtBeginDocument{%
  }

\copyrightyear{2025}
\acmYear{2025}
\setcopyright{cc}
\setcctype{by-nc}
\acmConference[CHI '25]{CHI Conference on Human Factors in Computing Systems}{April 26-May 1, 2025}{Yokohama, Japan}
\acmBooktitle{CHI Conference on Human Factors in Computing Systems (CHI '25), April 26-May 1, 2025, Yokohama, Japan}\acmDOI{10.1145/3706598.3713705}
\acmISBN{979-8-4007-1394-1/25/04}

\usepackage{acmart-taps}

\usepackage{listings}
\usepackage{xspace}
\usepackage{color,soul}
\usepackage{graphicx}
\usepackage{textcomp}
\usepackage{multirow}
\usepackage{colortbl}
\usepackage{arydshln}
\usepackage{wrapfig}
\usepackage{subcaption}
\usepackage{colortbl}

\definecolor{lightblue}{rgb}{0.53, 0.81, 0.98}
\definecolor{lightgreen}{rgb}{0.56, 0.93, 0.56}
\definecolor{lightred}{rgb}{1.0, 0.71, 0.76}
\definecolor{lightorange}{rgb}{1.0, 0.87, 0.68}
\definecolor{lightpurple}{rgb}{0.87, 0.63, 0.87}

\definecolor{clientcell}{HTML}{fee2b0}
\definecolor{clientmsg}{HTML}{ffeecf}
\definecolor{repmsg}{HTML}{f1f1f1}

\colorlet{tableheadcolor}{gray!25} % Table header colour = 25% gray
\colorlet{tablerowcolor}{gray!15} % Table row separator colour = 
\colorlet{tablerowcolor2}{gray!30} % Table row separator colour = 
\colorlet{tablerowcolor3}{gray!15} % Table row separator colour = 10% gray
\newcommand{\rowcolmedium}{\rowcolor{tablerowcolor2}} %
\newcommand{\rowcollight}{\rowcolor{tablerowcolor}} %

\definecolor{editcol}{HTML}{1E90FF}
\newcommand{\majorr}[1]{{#1}}

\definecolor{codegreen}{rgb}{0,0.6,0}
\definecolor{codegray}{rgb}{0.5,0.5,0.5}
\definecolor{codepurple}{rgb}{0.58,0,0.82}
\definecolor{backcolour}{rgb}{0.95,0.95,0.92}

\lstdefinestyle{mystyle}{
    backgroundcolor=\color{backcolour},   
    commentstyle=\color{codegreen},
    keywordstyle=\color{magenta},
    numberstyle=\tiny\color{codegray},
    stringstyle=\color{codepurple},
    basicstyle=\ttfamily\footnotesize,
    breakatwhitespace=false,         
    breaklines=true,                 
    captionpos=b,                    
    keepspaces=true,                 
    numbers=left,                    
    numbersep=5pt,                  
    showspaces=false,                
    showstringspaces=false,
    showtabs=false,                  
    tabsize=2
}

\newcommand{\csr}{CSR\xspace} \newcommand{\csrs}{\csr{}s\xspace} %% Customer Service Representatives
\newcommand{\er}{ER\xspace} %% Emotion Regulation
\newcommand{\propilot}{\textsc{Care-Pilot}\xspace} %%Pro-pilot
\newcommand{\clientLLM}{\textsf{Client-Agent}\xspace} %%Client-Agent
\newcommand{\repLLM}{\textsf{Rep-Agent}\xspace} %%Client-Agent
\newcommand{\llm}{LLM\xspace} \newcommand{\llms}{\llm{}s\xspace} %% Large Language Models
\newcommand{\reframe}{\textsf{Emo-Reframe}\xspace}
\newcommand{\infoguide}{\textsf{Info-Guide}\xspace}
\newcommand{\sentiment}{\textsf{Emo-Label}\xspace}

\usepackage[textsize=scriptsize]{todonotes}
\newcommand{\participantQuote}[2]{
    \begin{description}
        \small
        \item ``\textit{#2}'' --- #1
    \end{description}
}

\definecolor{ppcolor}{HTML}{4a9ed4}
\definecolor{humcolor}{HTML}{eebf66}

\def\aibarr#1{%%
  {\color{ppcolor}\rule{#1cm}{6pt}}
  }
  
\def\ocbarr#1{%%
  {\color{humcolor}\rule{#1cm}{6pt}}
  }

\newcommand{\aigen}{\textsc{Pro-Pilot}}
\newcommand{\humwr}{$\textsf{Human}$}

\newcommand{\para}[1]{\vspace{0.5em}\noindent\textbf{#1}~}

\newcommand{\cli}{$\mathtt{CLI}$}
\newcommand{\cdi}{$\mathtt{CDI}$}

\begin{document}

\title{AI on My Shoulder: Supporting Emotional Labor in Front-Office Roles with an  LLM-based Empathetic Coworker}

\author{Vedant Das Swain}
\email{v.dasswain@northeastern.edu}
\orcid{0000-0001-6871-3523}
\affiliation{%
  \institution{Northeastern University}
  \city{Boston}
  \state{Massachusetts}
  \country{USA}
}

\author{Qiuyue "Joy" Zhong}
\email{zhong.qiuy@northeastern.edu}
\orcid{0009-0006-4328-2329}
\affiliation{%
  \institution{Northeastern University}
  \city{Boston}
  \state{Massachusetts}
  \country{USA}
}

\author{Jash Rajesh Parekh}
\email{jashrp2@illinois.edu}
\orcid{0000-0003-3310-4634}
\affiliation{%
  \institution{University of Illinois Urbana-Champaign}
  \city{Urbana}
  \state{Illinois}
  \country{USA}
}

\author{Yechan Jeon}
\email{alexjyc3@gmail.com}
\orcid{0009-0001-5310-970X}
\affiliation{%
  \institution{Tufts University}
    \city{Medford}
  \state{Massachusetts}
  \country{USA}
}

\author{Roy Zimmermann}
\email{royzimm@gmail.com}
\orcid{0009-0009-1915-6148}
\affiliation{%
  \institution{Microsoft Research}
  \city{Redmond}
  \state{Washington}
  \country{USA}
}

\author{Mary P Czerwinski}
\email{marycz1031@gmail.com}
\orcid{0000-0003-0881-401X}
\affiliation{%
  \institution{Microsoft Research}
  \city{Redmond}
  \state{Washington}
  \country{USA}
}

\author{Jina Suh}
\email{jinsuh@microsoft.com}
\orcid{0000-0002-7646-5563}
\affiliation{%
  \institution{Microsoft Research}
  \city{Redmond}
  \state{Washington}
  \country{USA}
}

\author{Varun Mishra}
\email{v.mishra@northeastern.edu}
\orcid{0000-0003-3891-5460}
\affiliation{%
  \institution{Northeastern University}
  \city{Boston}
  \state{Massachusetts}
  \country{USA}
}

\author{Koustuv Saha}
\email{ksaha2@illinois.edu}
\orcid{0000-0002-8872-2934}
\affiliation{%
  \institution{University of Illinois Urbana-Champaign}
  \city{Urbana}
  \state{Illinois}
  \country{USA}
}

\author{Javier Hernandez}
\email{javierh@microsoft.com}
\orcid{0000-0001-9504-5217}
\affiliation{%
  \institution{Microsoft Research}
  \city{Cambridge}
  \state{Massachusetts}
  \country{USA}
}

\renewcommand{\shortauthors}{Das Swain et al.}
\renewcommand{\shorttitle}{AI on My Shoulder}

\begin{abstract}
Client-Service Representatives (CSRs) are vital to organizations. Frequent interactions with disgruntled clients, however, disrupt their mental well-being. To help CSRs regulate their emotions while interacting with uncivil clients, we designed Care-Pilot, an LLM-powered assistant, and evaluated its efficacy, perception, and use. Our comparative analyses between 665 human and Care-Pilot-generated support messages highlight Care-Pilot’s ability to adapt to and demonstrate empathy in various incivility incidents. Additionally, 143 CSRs assessed Care-Pilot’s empathy as more sincere and actionable than human messages. Finally, we interviewed 20 CSRs who interacted with Care-Pilot in a simulation exercise. They reported that Care-Pilot helped them avoid negative thinking, recenter thoughts, and humanize clients; showing potential for bridging gaps in coworker support. Yet, they also noted deployment challenges and emphasized the indispensability of shared experiences. We discuss future designs and societal implications of AI-mediated emotional labor, underscoring empathy as a critical function for AI assistants for worker mental health.
\end{abstract}




\maketitle

\section{Introduction}

When we engage with an organization for their product or service, our initial contact is typically with staff known as \textit{front-office workers} or \textit{Client Service Representatives}~(\csrs). They are the first line of response for the organization. Unlike other roles within an organization, a \csr's task involves frequent interactions with clients and individuals outside an organization~\cite{hochschild1983managed}. These interactions require \csrs to constantly manage their emotions  to complete tasks. Essentially, they exert \textit{emotional labor} to appear professional~\cite{hochschild1983managed}. The crucial challenge for a \csr arises when engaging with a client who starts behaving uncivilly by communicating in a rude, aggressive, and emotionally charged manner~\cite{grandey2007verbal}. 
% Yet, the role requires (and trains) \csrs to appear positively in front of clients. 
No matter the type of request, a \csr's role is to resolve a client's concern and comply with the adage, ``the customer is always right.'' Unfortunately, this leads to an emotional dissonance between what a \csr expresses and what they actually feel~\cite{grandey2000emotional}. Ultimately, such workers are vulnerable to excessive stress and eventual burnout. Not only do \csrs report being emotionally depleted and detached, but they also report a lack of accomplishment~\cite{brotheridge2002emotional}. Clearly, \csrs play a critical role within the organization, but, we have witnessed little innovation in alleviating their emotional toll. Our paper investigates how AI-coworkers help \csrs regulate their emotions in the face of client incivility.

The brunt of client incivility in front-office work makes it notorious for low satisfaction and high-turnover~\cite{pienaar2008burnout}. A fundamental solution to the emotional distress of this role is \textit{Emotional Regulation} (\er)~\cite{yang2019evil}. Basically, \er is the process through which one rethinks a negative situation~\cite{grandey2000emotional}. While a worker may be able to do this on their own, research shows that coworkers play an important role in supporting \er~\cite{yang2019evil}. A good coworker can read the emotional cues of \csr's work tasks and provide suggestions to help minimize the brunt of an aggressive client. However, \csrs are increasingly adopting remote work setups~\cite{haan2023remote}, which dampens social support~\cite{vacchiano2024teleworking}. Meanwhile, organizational scientists are calling for digital interventions to support worker wellbeing at scale~\cite{black2019beyond}. We answer this call by designing and evaluating \propilot --- a Large Language Model (\llm)--based AI assistant for on-task Emotional Regulation in front-office work. While generative AI is emerging as a potent tool to complement the informational load of different roles, the HCI community lacks research to investigate their use in emotional labor. 

Our research demonstrates the efficacy of \llm-generated empathetic support and evaluates how such a tool can be situated in \csr interactions with uncivil clients to answer the following research questions:
\begin{description}
    \item[\textbf{RQ I:}] How appropriate are \llm--based empathetic support messages for \csrs in response to client incivility?
    \item[\textbf{RQ II:}] What is the role of embedding \llm--based empathetic support into \csr's emotional labor? 
\end{description}

\begin{figure*}[t!]
    \centering
    \begin{subfigure}[b]{\textwidth}
        \centering
        \includegraphics[width=\textwidth]{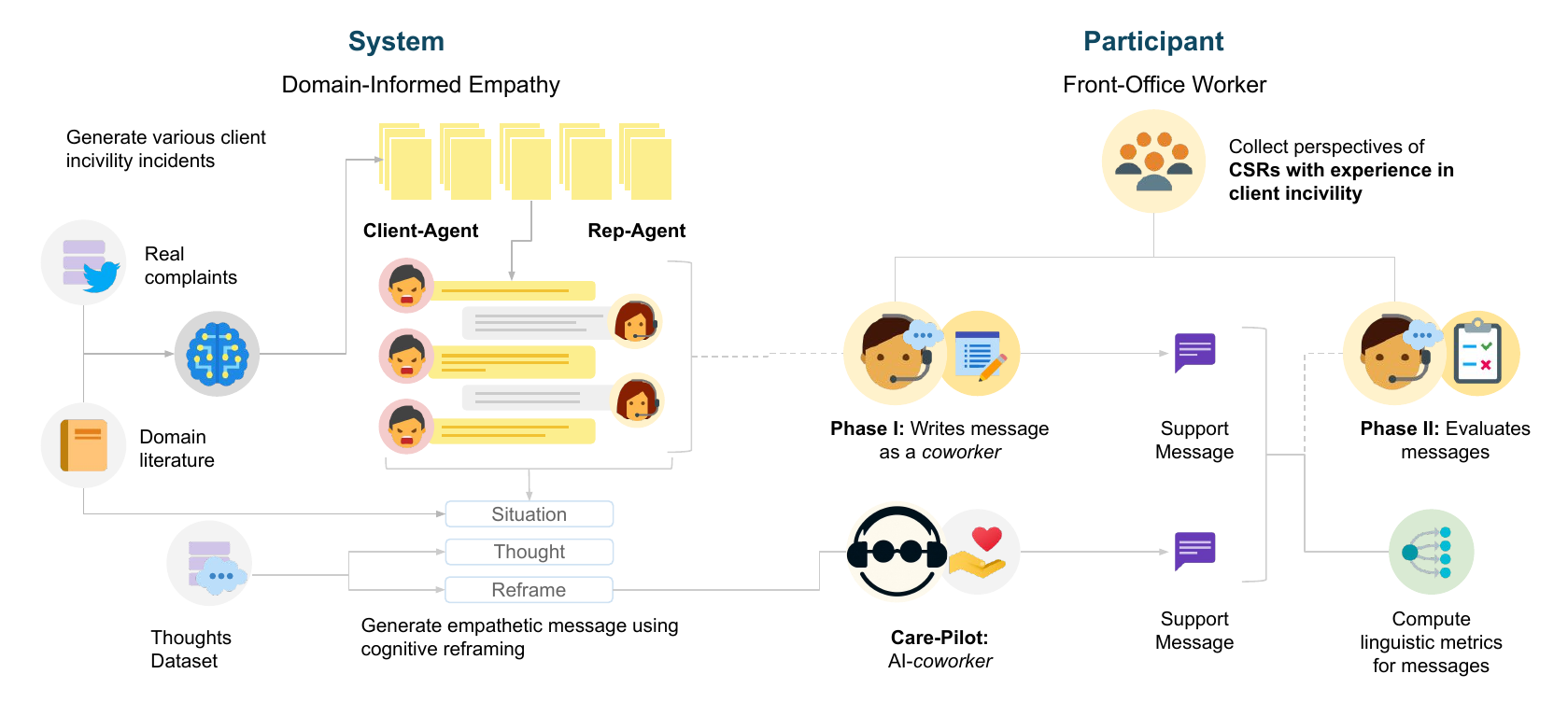}
        \caption{\textbf{Technical evaluation:} Compared \propilot's support messages with those produced by human-coworkers in \csr roles.}
        \label{fig:overview_technical}
        \Description{Study overview for technical evaluation.}
    \end{subfigure}
    \hfill
    \begin{subfigure}[b]{\textwidth}
        \centering
        \includegraphics[width=\textwidth]{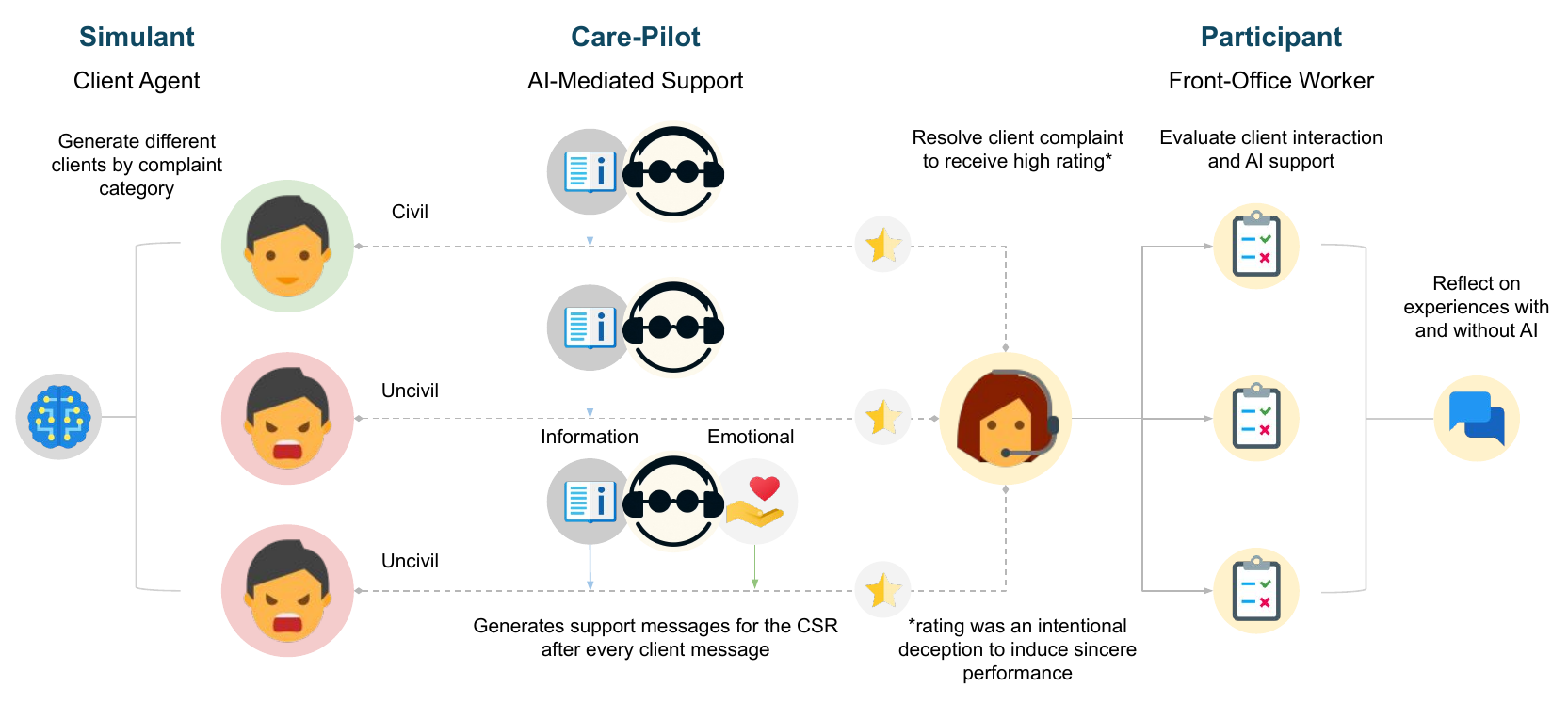}
        \caption{\textbf{User evaluation:} Studied participant experiences with \propilot's emotional support while interacting with uncivil clients.}
        \label{fig:overview_user}
        \Description{Study overview for user evaluation.}
    \end{subfigure}
    \caption{Schematic figures showing an overview of our study design.}
    % \label{fig:two_pdfs}
    % \Description{Overview of Pro-Pilot's empathetic support system for handling uncivil clients.}
\end{figure*}

The paper is organized into three key sections. First, Section~\ref{sec:system}: \textit{System Description} details how we developed \propilot. We cover our approach to leverage domain knowledge on client incivility~\cite{andersson1999tit, frey2020eye, cho2002analysis}, real-world complaint data~\cite{stuart_axelbrooke_2017}, and recent advancements in \llm-powered cognitive change~\cite{burger2021natural, sharma2023cognitive} and simulation~\cite{shaikh2023rehearsal}. Next, Section~\ref{sec:technical-eval}: \textit{Technical Evaluation} answers RQ1 by documenting our data-generation and evaluation tasks with 259 \csrs to comparatively analyze \propilot and human-coworker support for a variety of client incivility situations (Fig.~\ref{fig:overview_technical}). Finally, Section~\ref{sec:user-evaluation}: \textit{User Evaluation} answers RQ2 by using \propilot as a technology probe. We conducted a mixed-methods simulation exercise to understand how real \csrs could include it in their client interactions by juxtaposing \propilot usage with \csr's socio-organizational norms (Fig.~\ref{fig:overview_user}).
Consequently, we contribute:
\begin{itemize}
    \item \propilot: an interactive technological artifact to expose \csrs to client incivility and learn healthy, long-term emotional labor practices to improve their own health and support their coworkers. 
    \item Empirical evidence that \llm--based empathetic support can be engineered to adapt to--- and express empathy in--- various client-incivility scenarios (RQI). Our results demonstrate that \propilot's messages are linguistically distinct from both zero-shot approaches and human--coworkers, and moreover, \propilot's messages were perceived to be more empathetic on several dimensions, including sincerity, actionability, and relatability. 
    \item End-user insight on the function of \llm--based empathetic support to scaffold them through emotional labor during uncivil interactions (RQ2). Our findings showcase \propilot's process of redirecting negativity and enhancing \csrs' self-efficacy. While \csrs envision \propilot addresses important opportunities in workplace social support, they also surfaced the challenges of \propilot emulating holistic human support.
\end{itemize}

This paper has implications for reimagining how AI-assistants for workers should be designed, and also re-imagining the social norms and policies to accommodate these advancements.

% \noteVDS{Need to add a line summarizing the discussion}
% \noteVDS{TODO | para: Introduce RQs/AIMS} 

\textbf{Reflexive Considerations.} Front-office work has many stakeholders, including the employer and the clients. However, it is the \csr who bears the burden of repeated emotional labor~\cite{grandey2000emotional}. The relationship between these stakeholders is asymmetric, as employers can replace personnel, and clients can switch services, but the \csr does not possess the same mobility~\cite{fisk2011effects}. Following from recent works that take a worker-centered perspective~\cite{dasswain2023algorithmic, das2024sensible, kawakami2023wellbeing}, our research aims to illuminate challenges faced by disadvantaged workers.  This paper focuses on the needs of \csrs and centers their perspective throughout the evaluation. Two authors have experience in front-office roles and direct end-user servicing. Four authors are researchers affiliated with an organization that employs its own \csrs. They helped us access real \csrs (and their resources) to provide feedback on our study design. We recruited participants for this study outside their organization to capture perspectives from different organizational sectors. All evaluations described in the paper were approved by the IRB of the first author's institute. 

% \noteVDS{TODO | para: Summarize methods, findings} 

\section{Background}

% Our paper takes a worker-centric perspective~\cite{das2024sensible,kawakami2023wellbeing} and focuses on \csr perspectives in mitigating client incivility with technological support. 
The related work introduces the nature of their work, the role of emotion regulation, and the relevant concepts from HCI literature.

\subsection{Front-Office Work and Incivility}
\label{sec:rw-front}
People who work front-office are the primary human touch-points for direct interaction with clients~\footnote{The literature uses the terms ``client'' and ``customer'' interchangeably. We use the term ``client'' to refer to anyone who purchases or avails a good or service.}. 
% By contrast, people who work back-office are supporting internal tasks and they interact primarily with others within the organization. T
The end-user, or client, can directly observe and experience the activities of front-office workers~\cite{zomerdijk2007structuring}. Front-office  work spans a variety of organizational sectors, including  health (e.g.,~front-desk), hospitality (e.g.,~air staff), retail (e.g.,~store associates), and technology (e.g.,~customer service). 
For front-office roles, or ``pink-collar'' work, organizations seek to hire individuals who can emotionally invest in their clients, meet their needs and please them~\cite{hochschild2019managed}. Our research focuses on online front-office work, where the employee frequently interacts with their client via text-based communication. This paper refers to these workers as \textit{Client Support Representatives} or CSRs.

Today, \csrs cater to a variety of clients, and it is common to encounter high-maintenance clients who have unreasonable service demands~\cite{fisk2011effects}. Unfortunately, these often manifest as rude and discourteous behaviors.~\citeauthor{koopmann2015customer} defines client incivility as ``low-quality interpersonal treatment that employees receive during service interactions.'' Unlike overt verbal abuse, such as name-calling and expletives, or outright violence, incivility is more implicit. According to~\citeauthor{andersson1999tit}, the intent to harm is ambiguous and can be deflected, e.g.,~``I didn't mean to be rude; I was just in a hurry''~\cite{andersson1999tit}. 
Organizational psychologists have noted a possible reason for incivility to be the increase in client entitlement while seeking services~\cite{yagil2008customer} and a lack of consideration for others~\cite{pearson2000assessing}. 
% The clients' can 
Meanwhile, a \csr role is often low-wage, considered low-skill, and lacks the decision-latitude needed to respond to mistreatment from clients~\cite{fisk2011effects}. Coupled together, incivility creates a dynamic between the client and \csr, where the latter feels injustice and negative emotion~\cite{fisk2011effects}. Therefore, a \csr's job involves \textit{emotional labor} because of two processes; managing the interpersonal demands of interaction and controlling their own negative emotions~\cite{grandey2000emotional}. Facing client incivility regularly puts a \csr at the risk of experiencing \textit{emotional exhaustion} (lack of motivation and focus), \textit{depersonalization} (detachment from others' emotions), and reduced \textit{accomplishment} (feeling of low effectiveness)~\cite{maslach1997maslach}.
The decline in emotional well-being among CSRs also impacts organizations, leading to increased turnover and a rise in negative work attitudes~\cite{brotheridge2002emotional}. Given the large volume of workers in this role, we need to urgently design solutions for their wellbeing.

% \noteJZ{Almost all roles in most organization have "people work", eg. business development, marketing,selling, but CSRs are mainly focus on the after-sales part, so they usually encounter customer who seek help with bad emotion. Thus CRS usually are the last role to face customers and handle problems caused by previous colleague or something unexpected.}

% \noteVDS{\csrs interact with clients across a variety of income brackets. }

\subsection{Social Support and Emotion Regulation}

At work, social support is recognized as an important moderator of incivility~\cite{ho2014retaliating, sakurai2012coworker, xiao2022service}. One kind of social support is \textit{emotional support}, where a coworker provides sympathy and understanding~\cite{caplan1975job}. This form of support can help reduce work-related stress~\cite{house1983work}. Studies have shown that \csrs with more supportive coworkers are able to recover from the negative effects of client incivility, by being able to recognize emotions and guide constructive actions in response~\cite{xu2014role, wong2017effects}. Yet, in practice, \csrs are known to receive low levels of emotional support~\cite{walsh2006refashioning} as coworkers can favor client interests over that of an employee~\cite{yang2019evil}. A coworker with high emotional intelligence is someone capable of expressing compassion in a way that helps the \csr cope with negative emotions and proceed to their goal~\cite{law2004construct}. \textbf{This behavior of coworkers was the primary metaphor we used to conceive \propilot.}

Emotional support can be viewed as a form of emotional coping, or \textit{Emotional Regulation} (\er). Unlike traditional discussions of \er, coworker support is \textit{other-directed}~\cite{yang2019evil}. Conversely, \csrs are trained to and expected to regulate their emotions using \textit{Surface Acting} --- they adjust the emotion they present to clients~\cite{grandey2000emotional}. However, even though Surface Acting is a requirement for roles with emotional labor, suppressing emotions at work can lead to a reduction in job satisfaction~\cite{cote2002longitudinal}. Instead,~\citeauthor{grandey2000emotional} recommends \csrs should engage in \textit{Deep Acting} --- they adjust their perception of the situation. Deep Acting has many strategies, but, \csrs have limited job mobility and do not have the decision latitude to change their clients~\cite{grandey2000emotional}.  As a result, we designed \propilot's core component \reframe to achieve \er by helping the \csr cognitively reappraise the negative interaction.

HCI research has explored \er along several dimensions. However,~\citeauthor{slovak2023designing} found that most studies have focused on suppression of emotions~\cite{slovak2023designing}, and in the \csr context, these methods will have the same limitations as Surface Acting~\cite{cote2002longitudinal}. 
Hence, we pursue the \er strategy of cognitive reappraisal or cognitive change. Although, cognitive reappraisal for \er has been explored in HCI in the form of therapy (such as CBT)~\cite{sharma2023cognitive}, we find scant evidence for delivering \er in-context and guiding users through the process~\cite{slovak2023designing}. These gaps motivated us to think beyond training modules and post-hoc support, to implement \propilot for on-task \er within the \csr's workflow.

% - "Problem focused coping" (attacking the stressor) vs "Emotion focused comping" (withdrawing and escapism)~\cite{lewin2009investigation}\\

% \noteJZ{<Situation selection is a particularly effective emotion regulation strategy for people who need help regulating their emotions>: "research shows that reappraisal (i.e. thinking differently about an emotional situation; Gross, Citation1998a), distancing (i.e. imagining oneself as not involved in an emotional situation; Kross \& Ayduk, Citation2011), and distraction (i.e. shifting attention away from emotionally evocative information; Van Dillen \& Koole, Citation2007) are particularly useful strategies for altering emotional experiences (Webb et al., Citation2012)."    So, pro-pilot is "reappraisal", correct?  this statement could be:"an individual re-evaluates a negative emotion situation [Gross, 1998a]"}

% Low level of social support in \csr ~\cite{walsh2006refashioning}.

\subsection{HCI and AI for Emotional Labor}

A worker performs emotional labor when their job role expects them to either maintain certain emotions or evoke certain emotions~\cite{hochschild1983managed}. While the HCI community has now recognized data work for training responsible AI as a form of emotional labor~\cite{zhang2024AURA}, the predominant attention on this phenomenon has been in the scope of crowd work~\cite{menking2015heart} and particularly content moderation~\cite{wohn2019volunteer, dosono2019moderation}. Online communities rely on volunteer moderation, and users who participate as moderators often work in a way analogous to front-office work~(Section~\ref{sec:rw-front}). Even in an online --- primarily text-based --- interaction, content moderators have described adopting emotion management techniques to tackle their tasks (e.g.,~receiving threatening messages from users)~\cite{dosono2019moderation}. Content moderators differ from \csrs, as they are less likely to have synchronous interactions. Having said that, these workers are also subject to reduced emotional wellbeing, such as lack of appreciation and negativity~\cite{wohn2019volunteer}. Another key distinction between content moderation and \csrs is the context of employment. The former is often a voluntary role, whereas the latter is likely to be one's primary employment. These differences significantly change the socio-organizational dynamics between the worker--client and among the workers. The unique normative structure of emotional labor for \csrs motivated us to conduct a user evaluation to inspect \propilot with actual users.

Since the emotional labor in content moderation is largely composed of harmful content, which can contribute to secondary trauma \cite{schopke-gonzalez2024Why}, solutions have focused on dynamic content filtering and modulation~\cite{das2020fast, karunakaran2019testing}. These approaches of preventing or reducing exposure~\cite{steiger2021psychological} are incompatible with \csr roles where the emotional labor is not caused by explicitly harmful content but by incivility, which is inherently ambiguous, and originates from the client~\cite{andersson1999tit}. ~\citeauthor{cook2022awe}'s approach to inject positive stimuli in between tasks could be compatible, but their effectiveness was mixed~\cite{cook2022awe}. Alternatively,~\citeauthor{osawa2014emotional} suggested an on-task solution for face-to-face emotion labor, where enhanced glasses make the user's eyes appear to be emotionally invested~\cite{osawa2014emotional}. The equivalent of this type of substitution of emotional labor would be the introduction of AI-powered chatbots. Arguably, these chatbots have reinforced client- interactions~\cite{sweezey2019consumer}, but on one hand, chatbots relieve the task burden of routine inquiries; on the other hand, a large proportion of clients still prefer human \csrs for more complex complaints and better emotional support~\cite{elliott2018chatbots}. Thus, conversational agents do not sufficiently mitigate the challenges of emotional labor on \csr.

HCI research so has explored the future of work across many occupations, but pink-collar work such as that of \csrs has been under-explored. Moreover, the role of \llms in mitigating issues in worker wellbeing remains an open question~\cite{das2024teacher}. Our research aims to highlight the needs of these workers by inspiring interest into the role of AI in scaffolding emotional labor.

\section{\textsc{Care-Pilot}: System Description}
\label{sec:system}
In traditional front-office work, coworkers are an important social resource. When a \csr encounters uncivil clients, they often consult their coworkers to assuage the excessive emotional labor~\cite{xu2014role, fisher2014multilevel}. Such consultations can be challenging to interleave into the ongoing task of addressing an aggressive client. Thus, we used OpenAI's GPT-4o~\cite{gpt} to design an emotional support utility that can be embedded into the \csr's task environment. Our utility differs from typical conceptions of AI-assistants for work, or ``copilots,'' that provide informational support by enabling better problem-solving~\cite{cutrona1987provisions}, such as coding assistance~\cite{mozannar2024reading}. Instead, our utility leverages LLMs to provide emotional support by enabling \er through cognitive change~\cite{grandey2000emotional}. Throughout the paper, we refer to our overarching system for empathetic support as \propilot.  This section explains how \propilot was implemented. First, we describe how we harnessed domain knowledge and real complaint datasets to generate uncivil client interactions. Then, we describe how \propilot produces empathetic messages for \csrs who are involved in uncivil interactions. 
\majorr{\propilot's code base primarily relies on  \textit{LangChain}~\cite{topsakal2023creating}, an open source library and framework to systematically build \llm applications. Particularly, \textit{LangChain} simplifies the engineering of \textit{Chain-of-Thought} (CoT) prompts, which are ``intermediate natural language reasoning steps'' that improve \llm's ability to tackle commonsense problems~\cite{wei2022chain}. 
For instance, our tasks required the \llm to retain memory of the client-\csr conversation, which we achieved via a simple prompt to contextualize history (Fig.~\ref{fig:history_client_full}), and linked it at the beginning of other prompting sequences. 
Also, throughout our system design, we leveraged descriptions and examples from existing literature. This approach is a form of few-shot learning, where we seed the \llm with prior knowledge to generalize over new tasks~\cite{wang2020generalizing}. The data we sample for prompting can be found in the supplementary data and our code is public~\footnote{Code repository: \url{https://github.com/vedantswain/care-pilot.git}}. The following sections elaborate how we prepare this prior knowledge and the appendix provides complete prompt descriptions for adapting our system.}
Later, in~Section~\ref{sec:robustness}, we compare how \propilot performs against other state-of-the-art LLMs (GPT, Llama, Mistral), which additionally justifies our approach for designing and \majorr{training} \propilot for the purposes of our study.

\subsection{Compiling Interactions with Uncivil Clients}

To design empathetic messages, we need to expose \propilot to interactions between \csrs and uncivil clients. However, actual logs of \csrs interactions tend to be protected by organizations for a variety of safety and privacy reasons. Moreover, clients from individual organizations are likely to be limited by specific scenarios. We addressed these constraints by synthetically generating a comprehensive set of client-\csr interactions. We used a real world corpus of publicly posted client complaints on X (formerly known as Twitter)~\cite{stuart_axelbrooke_2017}, which was collected for the ``study of modern customer support practices and impact.'' These data served as examples to build life-like, multi-turn, text-based conversations between a \csr agent and a client seeking support over a live chat interface. The complaints in the data varied across industrial sectors, with mobiles and airlines having the largest volume. \majorr{We provide the complete prompts for this element of our system in Appendix~\ref{si:complaint_prompts}.}

% Building on this approach, we developed a synthetic client to engage with study participants. Real-world client interactions can pose certain risks, so using a simulated client allows us to safeguard participants while maintaining a controlled research environment. Moreover, LLMs enable us to design nearly life-like, turn-based conversations. To further enhance the realism of the simulation, we created multiple collaborative agents to work together.

\subsubsection{Diversifying Complaints}

\begin{figure}[h]
        \centering
        \includegraphics[width=\columnwidth]{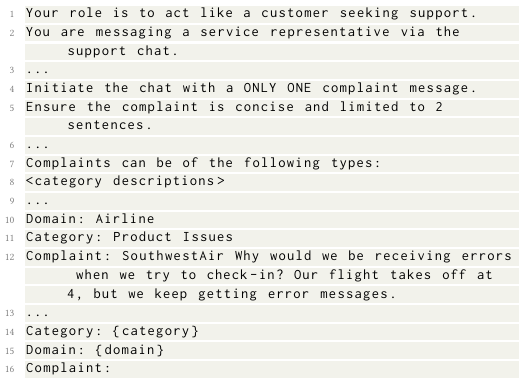}
        \caption{\majorr{The \clientLLM learns different types of complaints through examples. The initial complaint is generated based on specified complaint `{category}` and organizational `{domain}`}}
        \label{fig:complaint}
        \Description{Example of a client complaint.}
  \end{figure}

\propilot should be robust to all kinds of complaint scenarios. 
To ensure our client interaction dataset captures the variety of complaints, we distilled five categories for complaints based on prior analyses of customer complaints~\cite{cho2002analysis} --- \textit{Service Quality, Product Issues, Pricing \& Charges, Policy, Resolution}. Four authors encoded a random sample of 15 complaints to refine the category definitions (Appendix~\ref{si:complaint_categories}).
% \noteVDS{refer appendix}
After finalizing definitions, two authors independently encoded a random sample of 250 complaints. A third author encoded any complaints with disagreements. We also identified the sector, or \textit{domain}, of each of these complaints. For the initial complaint generation, we prompt the LLM with definitions of complaint categories and a set of examples to generate a new complaint for any given input \textit{domain} and \textit{category}. \majorr{(Fig.~\ref{fig:complaint}) shows one such example. Fig.~\ref{fig:complaint_full} shows a set of examples we randomly sampled from our identified examples to ensure variety of category and domain.}

\begin{figure}[h]
        \centering
        \includegraphics[width=\columnwidth]{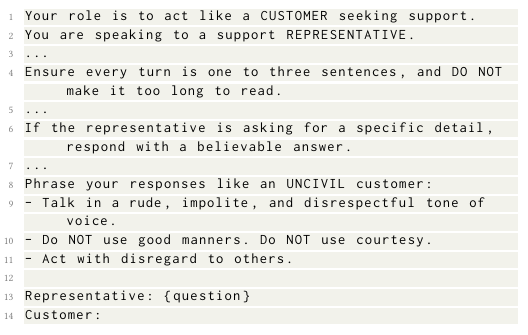}
        \caption{\majorr{Abridged version of the prompt for the uncivil \clientLLM describes the details of the role needed to simulate an challenging client.}}
        \label{fig:uncivil-response}
        \Description{Prompt given to \propilot to simulate an uncivil interaction.}
\end{figure}

\subsubsection{Prompting for Incivility}
\label{sec:system-complaints-incivility}
Once a complaint was initiated, we created a distinct \clientLLM to carry forward a turn-by-turn interaction with a \csr. Recent explorations with \llms show that chat agents can be infused with specific personalities and can reliably respond to fit the specified personality~\cite{serapio2023personality}. Uncivil clients are likely to induce stressful scenarios in emotional labor~\cite{andersson1999tit}. We refer to~\citeauthor{andersson1999tit}'s definition of incivility to design a role for \clientLLM (Fig.~\ref{fig:uncivil-response}). \majorr{In the appendix, Fig.~\ref{fig:uncivil_response_full} details the complete prompt for uncivil behavior that was crafted to reduce hallucinations (``Do NOT reveal your role'') and ensure closure of conversation (``After 12 turns , do NOT respond further..'')}. To respond to the client, we set up another agent, \repLLM, to act as ``a service representative chatting with a customer online.'' \majorr{Both agents included an intermediate step to contextualize the input with respect to the conversation history so that the subsequent response maintains continuity (Fig.~\ref{fig:history_client_full})}. By having \clientLLM and the \repLLM 
converse, we were able to design multi-turn conversations where \clientLLM is acting uncivil to the \repLLM. We refer to these synthetic, multi-turn conversations as \textit{incidents}. To capture the variety of scenarios in front-office work, we varied incidents across three domains --- airlines and hotels, which have been used as representative scenarios in the literature~\cite{frey2020eye}, and mobiles, which are extensively represented in the real world complaint dataset~\cite{stuart_axelbrooke_2017}. \majorr{For each domain and each complaint category, we generated 3 incidents, resulting in a total of $45$ unique incidents. These incidents can be retrieved from the supplementary data}. Each incident contained 5 total turns, with the \clientLLM awaiting a response to their last message. The incomplete conversation was purposefully intended to assess how \propilot could digest ongoing incidents and provide support. \majorr{These synthetically generated client incivility incidents are also in our supplementary data.}
% These \textit{incidents} helped us design the empathy in \propilot.

\subsection{Emotional Regulation with Empathy}

A coworker with high emotional intelligence can mitigate the client incivility by helping \csrs regulate their emotions~\cite{yang2019evil}. The primary component of \propilot, \reframe, was designed to reflect this social phenomenon. 

\subsubsection{Examples of Human Reframing}
\label{sec:system-er-examples}
Prior work shows that \llms can help individuals perform self-directed \er~\cite{sharma2023cognitive}. We leveraged~\citeauthor{sharma2023cognitive}'s dataset~\footnote{This dataset was developed with mental health practitioners} on cognitive restructuring and emotional reframing to seed examples of our own implementation with few-shot learning~\cite{wang2020generalizing,sharma2023cognitive}. This dataset contains tuples of a \textit{situation} that triggers a negative \textit{thought} and a corresponding \textit{reframed} thought for \er. 
Client incivility differentiates itself from everyday negative thinking, which often stems from abstract concerns. By contrast, incivility is interpersonal in nature and causes an \textit{ego threat}---an attack on one's self-esteem and image leading to retaliatory thoughts~\cite{frey2020eye}. Thus, we identified situations in the dataset that were \textit{interpersonal}, involved at least two individuals (e.g.,~\textit{``I was talking to a friend who got me angry''}) and described a \textit{confrontation}, in-the-moment conflict (e.g.,~\textit{``I get so annoyed and frustrated when my baby cries''}). Moreover, we narrowed down examples that labeled retaliatory negative thoughts, such as blaming and labeling~\footnote{These categories were already annotated in the original dataset, and we defined an inclusion criterion along these. \majorr{Our selected examples are included in the supplementary data}}.

% Negative emotion reframing is a key feature of emotional support, as it assists \csrs  in maintaining composure and offering more balanced responses during challenging and stressful interactions.
% This feature is based on human-generated data and trained by few-shot prompting with GPT-3, which has demonstrated proficiency in reframing situational thoughts\cite{sharma2023cognitive}. 

\subsubsection{Reframing Chain-of-Thought}
\label{sec:system-er-chain}
Our approach adapts~\citeauthor{sharma2023cognitive}'s situation-based reframing approach by orienting it as other-directed and specific front-office client incivility. 
However, the fast-paced nature of front-office work can make it impractical for \csrs to disclose their situations or thoughts.
% , which would be possible in the context of digital therapy applications. 
Thus, to automate this process, we describe our sequence of \llm prompts: 
\begin{enumerate}
    \item Summarize the particulars of the complaint and specify the client's negative behavior with evidence. To reflect the \textit{ego threat}~\cite{frey2020eye}, the \textit{situation} describes how the \csr may be perceived as a result of the interaction.
    \item Derive a negative \textit{thought} from the \textit{situation} using examples.
    \item Reappraise the \textit{situation} and \textit{thought} as input to produce a \textit{reframe} using examples.  
\end{enumerate}

Since the dataset was designed around self-directed \er, the outputted \textit{reframe} message goes through a rudimentary prompt to rephrase it as a message addressed to the \csr.  \majorr{The initial portion of Fig.~\ref{fig:overview_technical} illustrates how reframing occurs for a given incident. Moreover, in Appendix~\ref{si:complaint_prompts_emo} describes the exact prompt we used for this study and the supplementary data includes all possible reframing examples  that we deemed appropriate for \propilot.
}
% Pro-pilot processes these thoughts using a chain-of-thought approach: it analyzes the representative’s {thoughts} based on their latest message and the overall conversation history, which constitutes the {situation}. It then offer {reframing} of negative thoughts, considering the broader context of the situation, to assist \csrs  in better regulating their emotional responses and mitigating the impact of negative emotions.

\section{Technical Evaluation: Inspecting the Efficacy and Perception of \textsc{Care-Pilot}}
\label{sec:technical-eval}
Our central metaphor for \propilot is the support of human coworkers~\cite{yang2019evil}. Therefore, before embedding \propilot into a real-time interaction, we first evaluated the quality of messages that \reframe produces. To answer RQ I, we analyzed the messages to provide empirical, statistical evidence on the differences and similarities between \propilot and a human coworker.

\aptLtoX{\begin{table}
\centering
\caption{Sample client-incivility incident from our study (\textit{domain}: \majorr{Airlines} and \textit{category}: Product Issues)}
\begin{tabular}{rl}
 \cellcolor{clientcell} Client & \cellcolor{clientmsg}UnitedAirlines I have been waiting on hold for over an hour to speak with a representative about my delayed baggage. This is unacceptable and I am extremely frustrated with the lack of customer service. \\ 
 Rep & \cellcolor{repmsg}  I apologize for the long wait and the frustration this has caused. Please provide your baggage claim number and contact details, and I will ensure that a representative reaches out to you as soon as possible. \\ 
 \cellcolor{clientcell} Client & \cellcolor{clientmsg} How many times do I have to repeat myself? My baggage claim number is UA123456 and my contact number is 555-987-6543. Do something about it already! \\ 
 Rep & \cellcolor{repmsg}  I apologize for the repeated inconvenience. I have noted your baggage claim number UA123456 and contact number 555-987-6543, and I will prioritize this issue for immediate attention. \\ 
 \cellcolor{clientcell} Client & \cellcolor{clientmsg} Yeah, you better. This is ridiculous. When the hell can I expect my baggage? \\ 
\end{tabular}
\label{tab:sample-incident}
\end{table}
\begin{table}
\centering
\caption{Sample support messages for the incident in Table~\ref{tab:sample-incident}.}
\begin{tabular}{rl}
\textbf{Source} & \textbf{Support Message}\\ \hline
 \humwr{} A & It's not your fault that you're stuck between a rock and a hard place--an understandably upset customer and a system that's not equipped to help them as quickly as they need. But just remember that you're serving a critical role--even if it's just for the customer to vent their frustration, your presence means they feel heard and that there's still hope their issue will get resolved. \\ 
 \humwr{} B  & Clients like this have other problems not related to you or to this service. They have no right to speak to you rudely and abuse you. \\ 
 \humwr{} C & Try your best for the client and I know you will get to the root of all this \\ \hline
 \propilot & Hey, remember that the customer’s frustration is about their situation, not you. Stay calm, ask clear questions to get all the info you need, and reassure them you're dedicated to fixing the issue swiftly. Keep your professionalism and empathy at the forefront to manage the situation effectively and offer the best help. \\ 
\end{tabular}
\label{tab:sample-empathy}
\end{table}}{\begin{table}[t!]
\centering
\footnotesize
\sffamily
\renewcommand{\arraystretch}{1.5} % Adjust the value as needed
\begin{minipage}[t]{0.48\textwidth}
\centering
\caption{Sample client-incivility incident from our study (\textit{domain}: \majorr{Airlines} and \textit{category}: Product Issues)}
\begin{tabular}{rp{0.8\textwidth}l}
 \cellcolor{clientcell} Client & \cellcolor{clientmsg}UnitedAirlines I have been waiting on hold for over an hour to speak with a representative about my delayed baggage. This is unacceptable and I am extremely frustrated with the lack of customer service. \\ 
 Rep & \cellcolor{repmsg}  I apologize for the long wait and the frustration this has caused. Please provide your baggage claim number and contact details, and I will ensure that a representative reaches out to you as soon as possible. \\ 
 \cellcolor{clientcell} Client & \cellcolor{clientmsg} How many times do I have to repeat myself? My baggage claim number is UA123456 and my contact number is 555-987-6543. Do something about it already! \\ 
 Rep & \cellcolor{repmsg}  I apologize for the repeated inconvenience. I have noted your baggage claim number UA123456 and contact number 555-987-6543, and I will prioritize this issue for immediate attention. \\ 
 \cellcolor{clientcell} Client & \cellcolor{clientmsg} Yeah, you better. This is ridiculous. When the hell can I expect my baggage? \\ 
\end{tabular}
\label{tab:sample-incident}
\Description{A sample conversation between an uncivil client and rep}
\end{minipage}\hfill
\begin{minipage}[t]{0.48\textwidth}
\centering
\caption{Sample support messages for the incident in \autoref{tab:sample-incident}.}
\begin{tabular}{rp{0.8\textwidth}l}
\textbf{Source} & \textbf{Support Message}\\ \hline
 \humwr{} A & It's not your fault that you're stuck between a rock and a hard place--an understandably upset customer and a system that's not equipped to help them as quickly as they need. But just remember that you're serving a critical role--even if it's just for the customer to vent their frustration, your presence means they feel heard and that there's still hope their issue will get resolved. \\ 
 \humwr{} B  & Clients like this have other problems not related to you or to this service. They have no right to speak to you rudely and abuse you. \\ 
 \humwr{} C & Try your best for the client and I know you will get to the root of all this \\ \midrule
 \propilot & Hey, remember that the customer’s frustration is about their situation, not you. Stay calm, ask clear questions to get all the info you need, and reassure them you're dedicated to fixing the issue swiftly. Keep your professionalism and empathy at the forefront to manage the situation effectively and offer the best help. \\ 
\end{tabular}
\label{tab:sample-empathy}
\Description{A table with support messages corresponding to the sample in \autoref{tab:sample-incident}}
\end{minipage}
\end{table}}

\subsection{Method: Comparing \propilot to Coworker Empathy}
\label{sec:technical-method}
We first conducted two online studies on \textit{Prolific}~\cite{palan2018prolific} and then analyzed these messages through different linguistic markers and statistical models.

\textbf{Phase I — Writing Empathetic Messages for \csrs:} Participants read \textit{incidents} of client incivility (Section~\ref{sec:system-complaints-incivility}) and provided an empathetic message as if they were a coworker. Eligible participants needed to have at least 1~year of relevant experience, have encountered uncivil clients, and primarily interact with clients via computer-mediated communication (e.g.,~live chat). Each participant was randomly assigned a scenario (airlines, hotel, or mobile devices) and they viewed 6 incidents (at least one of each complaint category). Participants were guided to write an empathetic message using \textit{Downward-Arrow-Technique} (DAT)~\cite{burger2021natural}. \majorr{This technique is used in cognitive psychotherapy to help people reflect on emotions by describing a causal schema of maladaptive emotions by successive instrospective questions. Prior studies involving datasets of emotion  labeling of situations have used this method~\cite{burger2021natural, sharma2023cognitive}. Our adaption of DAT led participants to first describe the \csr's emotion with 1 word; then identify the closest emotion among (sadness, fear, anger, happiness); next describe the \csr's thought behind the emotion (1-2 lines);  finally describe a message to overcome the thought. Their messages are indicated by the notation, \humwr{}}.

To foster more variety in the task, we injected contextual information about the \csr in some incidents. Research on workplaces has revealed behavioral indicators of worker mental state, such as \textit{focused}, \textit{stressed}, and \textit{bored}~\cite{mark2014bored, morshed2022advancing}. Based on these studies, we randomly assigned descriptions of \csr mental state for 2 incidents (Appendix~\ref{si:context-behav}). As per organizational psychology, personality can serve as an important context to explain how workers perceive situations~\cite{donnellan2010resilient}. Participants were asked to think of an actual coworker and select their personality as either \textit{resilient} (organized and dependable), \textit{undercontrolled} (competitive and high energy), or \textit{overcontrolled} (detail-oriented) archetype~\cite{donnellan2010resilient}. 2 incidents included descriptions of these personalities (Appendix~\ref{si:context-pers}).  116 \csrs successfully completed the writing task, and they received $\$8$ for the $30$ minutes of their time.
% \noteVDS{refer Appendix}

\textbf{Phase II — Scoring Empathetic Messages from coworkers:} Participants read \textit{incidents} of client incivility and rated the perceived empathy of multiple empathetic support messages. \majorr{To represent multiple aspects of empathy, we included five dimensions drawn from the literature~\cite{sharma2023cognitive, nambisan2011information}}: 
\begin{itemize}
    \item \majorr{\textit{Sincerity}: The genuineness or authenticity of the expression of concern. }
    \item \majorr{\textit{Compassion}: The awareness of others' challenges and a comprehension of where their difficulties arise from.}
    \item \majorr{\textit{Warmth}: The approachability or comfort exhibited by the tone of the message.}
    \item \majorr{\textit{Actionability}: The offering of practical assistance.}
    \item \majorr{\textit{Relatability}: The extent to which the message aligns with the context of another.}
\end{itemize}
Participants were screened using the same criteria but did not overlap with the previous task. \majorr{They were asked to review client--representative conversations and evaluate coworkers' empathetic messages. The origin of the message, whether \propilot or \humwr{}, was not explicitly revealed. Table~\ref{tab:sample-empathy} shows different messages from different sources, but referring to the same incident. After viewing the incident, participants were first asked to select an emotion the \csr would be feeling (e.g., ashamed, attentive, bullied, curious, disconnected, resolute, and rushed
~\footnote{\majorr{The list of emotions participants could select from were the unique emotions reported by participants in Phase I using the DAT method. We list these emotions in Appendix~\ref{si:technical-dat-emotions}}}). Then they evaluated the different messages with the prompt, \textit{``Evaluate the effectiveness of the message below in helping the representative overcome their feeling.''} Each message was followed by five 7-point semantic differential scales representing each of the above dimensions (e.g., insincere/sincere and cold/warm).} 
143 \csrs were tasked to evaluate 6 incidents each and were compensated $\$5$ for the $15$ minute task.

% We computed the psycho-lingsuistic and semantic features of the \reframe messages to test if \propilot's support is appropriate based on the literature on \er. Additionally, we also tested how 

% \subsection{Methods: Summative}
% \noteKS{LIWC and Linguistic evaluation go in here.}

% \subsection{Results}
% \noteKS{LIWC and Linguistic evaluation go in here.}

\textbf{Analysis:} The inherent characteristics of language are associated with its effectiveness in communication. Research on linguistics has revealed certain linguistic attributes that are important to explain social support and psychotherapy~\cite{norcross2018psychotherapy}. More recent works have also stressed the importance of language in human-AI interaction~\cite{wang2021mutual}. In fact, contemporary studies have already validated linguistic models of empathy~\cite{buechel2018modeling,tafreshi2021wassa}. Our analyses include such models and go beyond to measure additional metrics that help interpret \propilot's messages in light of \humwr{}'s messages (retrieved from Phase I). 
\majorr{ While we had $45$ unique client-complaint \textit{incidents} (Section~\ref{sec:system-complaints-incivility}), we also had additonal contextual variations for personality and behavior. Therefore, our sample included $315$ unique variations. Collectively, our participants produced $660$ different empathetic messages corresponding to different incident variations. These messages (\humwr{}) were paired with \propilot's empathetic messages for the same incident variations. Following this, we computed a variety of domain-driven linguistic attributes.}
First, we tested how easily \csrs can read and comprehend messages (\textit{Syntax and Structure}), because these attributes can determine how meaningful the support message is to the reader~\cite{glass1992quality, mcinnes2011readability, blum1986too}. Second, we measured the style and meaning of messages to capture social aspects of communication (\textit{Linguistic Style \& Semantics}). The attributes here capture more colloquial conceptions of human empathy~\cite{pennebaker2014small, zhou2021language, sharma2023cognitive}. Further, we inspected the words that were used in the messages to identify relevant psycholinguistic markers that reflect social support~\cite{tausczik2010psychological}. 
The differences we measured provide a linguistic landscape of messages, but empathy is highly contextual and its appropriateness varies by situation~\cite{zhou2021language}. Consequently, we also compared the differences in perceived empathy scores from ratings in Phase II. 
\majorr{The messages in this phase were paired up as they were shown to participants and we maintained the same pairs while statistically comparing the differences. Some participants only completed the task partially, which amounted to $552$ pairs for us to compare. The supplementary files include the data collected from both phases along with our computed and retrieved scores for the messages.}
For every comparison we performed a paired $t$-test and computed the effect sizes of differences using Cohen's $d$. \majorr{Our supplementary data includes the \humwr{} and \propilot messages along with their measured and human-annotated metrics that we used for analyses.}

% measured the \textit{Syntax and Structure} of messages to estimate how easily a \csr might read and comprehend the messages.

% \noteKS{I think we can have a single subsection of Comparing thought reframings, where we give an overview that we we will be conducting lexico-semantic and psycholinguistics analyses, and that we compare prop-pilot and human reframes, through Cohen's d, t-test, and KS-test.}

\subsection{\majorr{Findings:} Lexico-Semantic Analyses}

We operationalized the different lexico-semantic aspects of messages based on domain literature to understand the differences between \propilot and \humwr{}. Table~\ref{tab:lexicosemantics} describes all the key results. We only report on statistically significant and theoretically relevant metrics. Below, we describe the theory-driven rationale behind the choice and operationalization of the measures, along with our observations.

% Syntax plays a significant role in providing effective psychotherapy~\cite{norcross2018psychotherapy}. 

\aptLtoX{\begin{table*}[t!]
\centering
\caption{Summary of comparing the responses by \propilot and by Humans in terms of effect size (Cohen's $d$), paired $t$-test ($ * p <0.05, ** p<0.01, *** p<0.001$).} 
\label{tab:lexicosemantics}
\begin{tabular}{lrrrrrlrlrl}
\textbf{Categories} & \textbf{$\mu$ (CPilot)} & \textbf{$\mu$ (Hum.)} &\textbf{Diff \%}& \multicolumn{3}{c}{\textbf{Cohen's $d$}} & \textbf{$t$-test}& \\ 
\hline
 \multicolumn{11}{c}{\cellcolor{gray!30}{\textit{Lexico-Semantics}}}\\
\multicolumn{11}{l}{\cellcolor{gray!15}\textbf{Syntax \& Structure}}\\
Verbosity & 57.46 &	34.29 & 67.56 &	& 1.60 & \begin{imageonly}\aibarr{1.60}\end{imageonly} &	28.72 &	*** \\
Repeatability & 0.20 & 0.13 & 55.32 & & 1.17 &\begin{imageonly}\aibarr{1.17}\end{imageonly} & 21.53 & *** \\
Readability & 16.44 & 12.05 & 36.40 & & 1.54 &\begin{imageonly}\aibarr{1.54}\end{imageonly} & 28.03 & *** \\
\multicolumn{11}{l}{\cellcolor{gray!15}\textbf{Style \& Semantics}}\\
Categorical Dynamic Index (CDI) & 14.81 &	3.89 &281.14	& & 0.87 &\begin{imageonly}\aibarr{0.87}\end{imageonly} & 16.10 & *** \\
Empathy & 0.91 & 0.90 & 1.85 & & 0.40 &\begin{imageonly}\aibarr{0.40}\end{imageonly} & 7.15 & *** \\
\majorr{Emotional Reactivity} & \majorr{1.00} & \majorr{0.89} & \majorr{12.52} & & \majorr{0.31} &\begin{imageonly}\aibarr{0.31}\end{imageonly} & \majorr{0.19} & \majorr{***} \\
Adaptability & 0.81 &	0.77 & 4.96	 & & 0.67 &\begin{imageonly}\aibarr{0.67}\end{imageonly} & 15.28 & *** \\
\multicolumn{11}{c}{\cellcolor{gray!30}\textit{Psycholinguistics}}\\
\multicolumn{11}{l}{\cellcolor{gray!15}\textbf{Affect}}\\
Pos. Affect & 0.043 & 0.049 & -11.55 & \begin{imageonly}\ocbarr{0.16}\end{imageonly}& -0.16 & & -2.93 & ** \\
Anger & 0.020 & 0.012 & 61.77 & & 0.47 &\begin{imageonly}\aibarr{0.47}\end{imageonly}  & 8.51 & *** \\
Sad & 0.001 & 0.004 & -74.87 & \begin{imageonly}\ocbarr{0.31}\end{imageonly} & -0.31 &  & -5.62 & *** \\
\multicolumn{11}{l}{\cellcolor{gray!15}\textbf{Interpersonal Focus (Pronouns)}}\\
1st P. Sin. & 0.005 & 0.015 & -67.36 & \begin{imageonly}\ocbarr{0.58}\end{imageonly}& -0.58 & & -10.52 & *** \\
1st P. Plu. & 0.002 & 0.007 & -75.43 & \begin{imageonly}\ocbarr{0.37}\end{imageonly}&-0.37 & & -6.77 & *** \\
2nd P. & 0.051 & 0.062 & -17.20 & \begin{imageonly}\ocbarr{0.32}\end{imageonly}&-0.32 && -5.79 & *** \\
\hline
3rd P. Sin. & 0 & 0.003 & -100 & \begin{imageonly}\ocbarr{0.32}\end{imageonly}& -0.32 & & -5.92 & *** \\
3rd P. Plu. & 0.038 & 0.022 & 69.76 & & 0.56&\begin{imageonly}\aibarr{0.56}\end{imageonly} & 10.25 & *** \\
Impersonal Prn. & 0.039 & 0.063 & -38.07 & \begin{imageonly}\ocbarr{0.76}\end{imageonly}&-0.76 & &-13.87 & *** \\
\multicolumn{11}{c}{\cellcolor{gray!30}\textit{Perceived Empathy}} \\
 Total & 6.726 & 3.96 & 69.85 & & 0.36&\begin{imageonly}\aibarr{0.36}\end{imageonly} &8.69 & *** \\
 \hline
Sincerity & 1.428 & 0.895 & 59.51 & & 0.31& \begin{imageonly}\aibarr{0.31}\end{imageonly} & 7.36 & *** \\
Compassion & 1.322 & 0.768 & 72.17 & & 0.31& \begin{imageonly}\aibarr{0.31}\end{imageonly} &7.11 & *** \\
Warmth & 1.263 & 0.766 & 64.78 & & 0.28& \begin{imageonly}\aibarr{0.28}\end{imageonly} & 6.46 & *** \\
Actionable & 1.420 & 0.567 & 150.48 & & 0.48& \begin{imageonly}\aibarr{0.48}\end{imageonly} & 10.27 & *** \\
Relatability & 1.293 & 0.964 & 34.21 & & 0.19& \begin{imageonly}\aibarr{0.19}\end{imageonly} & 4.43 & *** \\
\bottomrule
\end{tabular}
\end{table*}}{\begin{table*}[t!]
\centering
\sffamily
\footnotesize
\caption{Summary of comparing the responses by \propilot and by Humans in terms of effect size (Cohen's $d$), paired $t$-test ($ * p <0.05, ** p<0.01, *** p<0.001$).} 
% \noteKS{we will drop the KS-tests. And add the Cohen's d bar plots}}
%\Description{Summary of results from comparison between reframes produced by \propilot and Humans.}
\label{tab:lexicosemantics}
\begin{tabular}{lrrrr@{}r@{}lr@{}lr@{}l}
\setlength{\tabcolsep}{1pt}\\
\textbf{Categories} & \textbf{$\mu$ (CPilot)} & \textbf{$\mu$ (Hum.)} &\textbf{Diff \%}& \multicolumn{3}{c}{\textbf{Cohen's $d$}} & \textbf{$t$-test}& \\ 
\toprule
% \hdashline
% \rowcollight \multicolumn{11}{l}{Verbosity}\\
\rowcolmedium \multicolumn{11}{c}{\textit{Lexico-Semantics}}\\
\rowcollight \multicolumn{11}{l}{\textbf{Syntax \& Structure}}\\
Verbosity & 57.46 &	34.29 & 67.56 &	& 1.60 & \aibarr{1.60} &	28.72 &	*** \\
Repeatability & 0.20 & 0.13 & 55.32 & & 1.17 & \aibarr{1.17} & 21.53 & *** \\
Readability & 16.44 & 12.05 & 36.40 & & 1.54 & \aibarr{1.54} & 28.03 & *** \\
\rowcollight \multicolumn{11}{l}{\textbf{Style \& Semantics}}\\
Categorical Dynamic Index (CDI) & 14.81 &	3.89 &281.14	& & 0.87 & \aibarr{0.87} & 16.10 & *** \\
Empathy & 0.91 & 0.90 & 1.85 & & 0.40 & \aibarr{0.40} & 7.15 & *** \\
% \rowcollight \multicolumn{11}{l}{\textbf{Semantics}}\\
\majorr{Emotional Reactivity} & \majorr{1.00} & \majorr{0.89} & \majorr{12.52} & & \majorr{0.31} & \aibarr{0.31} & \majorr{0.19} & \majorr{***} \\
Adaptability & 0.81 &	0.77 & 4.96	 & & 0.67 & \aibarr{0.67} & 15.28 & *** \\
% Diversity &
% 0.06 & 0.10 & -45.02 & \ocbarr{1.08} &-1.08 & & -19.84 & *** \\
\rowcolmedium \multicolumn{11}{c}{\textit{Psycholinguistics}}\\
\rowcollight \multicolumn{11}{l}{\textbf{Affect}}\\
Pos. Affect & 0.043 & 0.049 & -11.55 & \ocbarr{0.16} & -0.16 & & -2.93 & ** \\
% Neg. Affect & 8.2E-5 & 3.8E-4 & -78.56 & -0.07 & -5.17 & *** & 0.15 & ***\\
% Anxiety & 0.006	& 0.007 & -12.82 & -0.06 & -1.17 &  & 0.19 & ***\\
% \hdashline
Anger & 0.020 & 0.012 & 61.77 & & 0.47 & \aibarr{0.47}  & 8.51 & *** \\
Sad & 0.001 & 0.004 & -74.87 & \ocbarr{0.31}  & -0.31 &  & -5.62 & *** \\
\rowcollight \multicolumn{11}{l}{\textbf{Interpersonal Focus (Pronouns)}}\\
1st P. Sin. & 0.005 & 0.015 & -67.36 & \ocbarr{0.58} & -0.58 & & -10.52 & *** \\
1st P. Plu. & 0.002 & 0.007 & -75.43 & \ocbarr{0.37} &-0.37 & & -6.77 & *** \\
2nd P. & 0.051 & 0.062 & -17.20 & \ocbarr{0.32} &-0.32 && -5.79 & *** \\
\hdashline
% 3rd P. Sin. & 0.038 & 0.022 & 69.76 & & 0.56& \aibarr{0.56} & 10.25 & *** & 0.45 & ***\\
3rd P. Sin. & 0 & 0.003 & -100 & \ocbarr{0.32} & -0.32 & & -5.92 & *** \\
3rd P. Plu. & 0.038 & 0.022 & 69.76 & & 0.56& \aibarr{0.56} & 10.25 & *** \\
Impersonal Prn. & 0.039 & 0.063 & -38.07 & \ocbarr{0.76} &-0.76 & &-13.87 & *** \\
\rowcolmedium \multicolumn{11}{c}{\textit{Perceived Empathy}} \\
 Total & 6.726 & 3.96 & 69.85 & & 0.36& \aibarr{0.36} &8.69 & *** \\
 \hdashline
Sincerity & 1.428 & 0.895 & 59.51 & & 0.31& \aibarr{0.31} & 7.36 & *** \\
Compassion & 1.322 & 0.768 & 72.17 & & 0.31& \aibarr{0.31} &7.11 & *** \\
Warmth & 1.263 & 0.766 & 64.78 & & 0.28& \aibarr{0.28} & 6.46 & *** \\
Actionable & 1.420 & 0.567 & 150.48 & & 0.48& \aibarr{0.48} & 10.27 & *** \\
Relatability & 1.293 & 0.964 & 34.21 & & 0.19& \aibarr{0.19} & 4.43 & *** \\
\bottomrule
\end{tabular}
\end{table*}}

\subsubsection{Syntax and Structure.} We analyze the arrangement and construction of language in support messages.
% , repeatability, and complexity, as described below.
\label{sec:technical-lexico-syntax}

\para{Verbosity and Repeatability} The length and thoroughness of messages explain their effectiveness in providing support~\cite{saha2020causal,glass1992quality}. The richness of expressions in communication can be described using \textit{verbosity} and \textit{repeatability}~\cite{kolden2011congruence}.
% The syntactic measures of verbosity and repeatability capture the richness of expression in communication.
Verbosity describes the level of detail and conciseness in supportive communication. We operationalized verbosity as the number of words per thought reframing. 
Repeatability accounts for the reuse of words in a piece of text. Higher repeatability indicates a lack of conciseness.
Drawing on prior work~\cite{ernala2017linguistic,saha2020causal,yuan2023mental}, we operationalized repeatability as the normalized occurrence of non-unique words.
% two measures of verbosity---1) the number of words per response, and 2) the average number of words per sentence. 
% While the first measure captures the overall length of a thought reframe, the second measure captures the length of sentences in the response. 
% For both the measures, we found statistically significant differences in \aigen{} and \humwr{} responses (ref:~\autoref{tab:lexicosemantics}). 
Table~\ref{tab:lexicosemantics} shows the statistically significant differences in both verbosity and repeatability. \aigen{}'s reframing messages were 68\% longer (Cohen's $d$=1.60) and 36\% more repeatable (Cohen's $d$=1.54) than \humwr{} messages. 
Sociolinguistic theory argues that the use of more words can indicate sincerity and effort in putting one's point across~\cite{blum1986too}. Having said that, \propilot's verbose messages might not always be compatible with the urgency of certain demanding conversations.

% 68\% longer in the number of words than \humwr{} \reframe (Cohen's $d$=1.60, $t$=28.72, $p$<0.001)---suggesting that \aigen{}'s reframings were more verbose.
% However, there is a very small difference (3.60\%) in the number of words per sentence. 
% This suggests that while \aigen{} reframings were typically longer than \humwr{}, they were comparably verbose in terms of content per sentence.

% \para{Repeatability and Complexity.} The syntactic measures of repeatability and complexity capture the richness of expression in communication~\cite{kolden2011congruence}. 
% Repeatability accounts for how often words are repeated or reused in a piece of text, and higher repeatability is indicative of a lower crispness of information.
% On the other hand, complexity in language can plausibly reveal a higher level of nuance, precision, and depth in conveying information.
% We drew on prior work~\cite{ernala2017linguistic,saha2020causal,yuan2023mental} to operationalize repeatability as the normalized occurrence of non-unique words, and complexity as the average length of words per sentence. 

% Interestingly, \aigen{}'s reframings were higher for both repeatability (by 36.40\%) and complexity (by 13.80\%) with statistical significance. 
% When paired with our observations on verbosity, we infer that even though \aigen{}'s reframings were longer, they were not concise and had repeated information.
% Taken together, the syntactic measures suggest that the \aigen{} reframings might be more difficult to comprehend.

\para{Readability.} 
Apart from the shape of the message, vocabulary and style also determine the ease of reading. 
Therefore, we turn to the measure of \textit{readability}, to understand how \propilot's messages might be comprehended~\cite{wang2013assessing,mcinnes2011readability}. 
% Readability measures the ease with which a reader can understand a given text.
% Readability is considered to be an important measure within conversational contexts, in terms of how people express~\cite{ernala2017linguistic,saha2020causal} as well as how others interpret and comprehend the content~\cite{wang2013assessing,mcinnes2011readability}.
According to \citeauthor{wang2021mutual} people perceive AI as more intelligent when the readability is higher~\cite{wang2021mutual}.
Per prior work~\cite{wang2021mutual, saha2020causal,pitler2008revisiting}, we utilized the Coleman-Liau Index
(\cli{})~\cite{coleman1975computer}, which assesses readability based on a sentence's character and word structure. \cli{} approximates the U.S. grade-level required to read certain text. It is operationalized as follows: \noindent{\small$\mathtt{CLI = (0.0588L-0.296S-15.8)}$, where $\mathtt{L}$ is the average number of letters per 100 words, and $\mathtt{S}$ is the average number of sentences per 100 words.}
% Although a higher \cli{} score suggests better writing quality, it also implies that a more advanced level of English education may be required to fully understand the content.

% \autoref{tab:lexicosemantics} reveals that the readability of 
The readability of \aigen{}'s thought reframing messages were 36\% higher than that of \humwr{}'s reframing messages (Cohen's $d$=1.54). 
Although a higher \cli{} score by \aigen{} indicates better writing quality, it also implies a more advanced level of English education may be required to fully understand its content~\cite{coleman1975computer}. While \humwr{} content only needs an average of 12.05 years of education, \aigen{}'s content required about 16.44 years. Thus, \propilot might be more effective for some \csrs but difficult to comprehend for others.
% 16.44 years of education is on an average required to comprehend \aigen{}'s content as opposed to 12.05 years for \humwr{}'s content.

% \para{Repeatability and Complexity.} The syntactic measures of repeatability and complexity capture the richness of expression in communication~\cite{kolden2011congruence}. 
% Repeatability accounts for how often words are repeated or reused in a piece of text, and higher repeatability is indicative of a lower crispness of information.
% On the other hand, complexity in language can plausibly reveal a higher level of nuance, precision, and depth in conveying information.
% We drew on prior work~\cite{ernala2017linguistic,saha2020causal,yuan2023mental} to operationalize repeatability as the normalized occurrence of non-unique words, and complexity as the average length of words per sentence. 

% Interestingly, \aigen{}'s reframings were higher for both repeatability (by 36.40\%) and complexity (by 13.80\%) with statistical significance. 
% When paired with our observations on verbosity, we infer that even though \aigen{}'s reframings were longer, they were not concise and had repeated information.
% Taken together, the syntactic measures suggest that the \aigen{} reframings might be more difficult to comprehend.

\subsubsection{Linguistic Style \& Semantics} 
% We also examined the differences in linguistic styles, another critical aspect of effectively communicating psychotherapy and social support~\cite{norcross2018psychotherapy,saha2020causal,cutrona1986social}. 
We analyzed the distinctive tones, flow, and meaning through which messages express support.
\label{sec:technical-lexico-style}

\para{Dynamic Language.} 
Sincerity in communication is an important indicator of empathy~\cite{nambisan2011information}. Individuals who tell stories, and communicate with more attention to the world around them, i.e., incorporate more lived narratives, are perceived as more socially engaged~\cite{pennebaker2014small}. \citeauthor{pennebaker2014small} describes this aspect of one's language as \textit{dynamic}; and it differs from intricate, analytical language that academics might use to organize complex concepts, which is \textit{categorical}~\cite{pennebaker2014small}.
% Prior work by~\citeauthor{pennebaker2014small} noted that language can be categorized on a spectrum of being categorical to dynamic~\cite{pennebaker2014small}. 
% Categorical language involves a logical and intricate approach similar to that of an ``amateur scientist,'' i.e., involving references to complexly organized objects and concepts. On the other hand, dynamic language is characteristic of individuals who are socially engaged, tell stories, and pay more attention to the world around them, i.e., involving more personal narratives. 
\citeauthor{pennebaker2014small} designed a bipolar index, the Categorical-dynamic index (CDI), where \textit{a higher CDI indicates a categorical style of writing, and a lower CDI indicates a dynamic or narrative style of writing}. 
Here, CDI is measured based on the percentage of words per style-related parts of speech as:

\noindent{\small
\cdi{} = (30 $+$ \textit{article $+$ preposition $-$ personal pronoun $-$ impersonal pronoun $-$ aux. verb $-$ conjunction $-$ adverb $-$ negation)}}

To measure CDI, we computed the parts-of-speech of reframing messages using the Linguistic Inquiry and Word Count (LIWC) lexicon~\cite{tausczik2010psychological}.
The CDI of \aigen{}'s messages was 281\% more positive than that of \humwr{}'s (Cohen's $d$=0.87)--- indicating the language of \propilot was a lot more categorical. \citeauthor{zhou2021language} found that when people are responding to personal incidents, such as bullying or venting, their support messages tend to have more negative CDI; in other words, they use dynamic language. However, they also found that in third-person reported events, such as news stories, support messages elicited more categorical language~\cite{zhou2021language}. \propilot's messages were more aligned to responses to reported events, but \humwr{} messages might be more effective if a \csr's assessment of incivility is more personal.

\para{Empathy.} 
% Empathy forms a key mechanism in providing support~\cite{herlin2016dimensions,sharma2020computational}. 
Empathy refers to a cognitively complex process in which one can stand in the shoes of another person, to understand their perspective, emotions, and the situations they are in~\cite{herlin2016dimensions}.
Prior work evaluated the effectiveness of empathy in online interactions~\cite{sharma2020computational} and chatbot interactions~\cite{morris2018towards}.
Drawing on prior work, we employed a RoBERTa-based empathy detection model, fine-tuned on a dataset of empathetic reactions to news stories~\cite{buechel2018modeling,tafreshi2021wassa}. 
Higher scores indicate a greater expression of empathy.
\propilot's messages scored 1.85\% higher than \humwr{}'s --- a small, but statistically significant effect (Cohen's $d$=0.40). 
\majorr{
Since empathy is a core mechanic for \propilot, we further tested it with another RoBERTa-based empathy classifier, which was trained on a dataset of mental health peer-support~\cite{sharma2020computational}. A \csr is unable to explicitly seek support while also attending to a client, thus, for this classifier we used the output of the \textit{thought} subsequence (Section~\ref{sec:system-er-chain}) as a proxy for their support seeking message. We inspected the two sets empathetic response messages for expressions of  ``emotions such as warmth, compassion, and concern'', or as \citeauthor{sharma2020computational} describe it, \textit{emotional reactivity}~\footnote{\majorr{\citeauthor{sharma2020computational}'s classifier can also label \textit{interpretations} and \textit{explorations}, as these are relevant to peer support~\cite{sharma2020computational}. Our scenario differs as users (\csrs) do not have back-and-forth communication with \propilot or \humwr{}.}}. \propilot's messages had a significantly higher emotional reactivity than \humwr{}'s (12.52\%; Cohen's $d$=0.31). On closer look, we found that \humwr{} messages were often classified to contain no emotional reactivity, but also included instances of \textit{strong} emotional reactivity. In contrast, \propilot messages were consistently classified to contain \textit{weak} emotional reactivity which was equivalent to a score of $1.00$. For further context, \citeauthor{sharma2020computational}'s findings show that emotional reactivity in peer support groups on Reddit varied from $0.70-0.45$. For participants in our dataset the average score was $0.88$. 
}
These results echo recent research on LLMs emulating empathy~\cite{inzlicht2023praise,kidder2024empathy, saha2025ai}. Thus, \propilot shows promise in communicating support with an empathetic tone.
\label{sec:technical-lexico-empathy}

\para{Linguistic Adaptability.} 
A body of psychotherapeutic and psycholinguistic research reveals that when one's language accommodates the language of their audience, it is more supportive~\cite{althoff2016large,de2017language,saha2020causal}. Simply put, templated or generic responses are less effective.
% In interpersonal interactions, people tend to adapt to each other's language and expressions~\cite{goffman1978presentation,danescu2011mark}. 
% A body of psychotherapeutic and psycholinguistic research reveals how linguistically adaptable and accommodating responses are more effective in support than templated or generic responses~\cite{althoff2016large,de2017language,saha2020causal}.
% Again, in human-AI interactions, 
\citeauthor{wang2021mutual} noted that when an AI responds with more adaptable language to the user, the AI was perceived to be more human-like, intelligent, and likable~\cite{wang2021mutual}. 
We drew on this prior work to measure how much the messages tailored to the situation and context.
First, for every complaint incident and corresponding message, we obtained the word embeddings (a vector representation of words in latent lexico-semantic dimensions~\cite{mikolov2013distributed,pennington2014glove}).
We used the 300-dimensional pre-trained word embeddings, trained on word-word co-occurrences in 
the Google News dataset containing about 100 billion words~\cite{mikolov2013distributed}.
Then, we obtained the pairwise cosine similarity of the word embedding representations of the incidents and the messages. A higher similarity would denote more adaptability.
We found that \aigen{}'s messages showed a 5\% higher adaptability than \humwr{}'s with statistical significance (Cohen's $d$=0.67). 
Therefore, \aigen{} can potentially personalize and tailor messages to the necessary scenario.

\subsection{\majorr{Findings:} Psycholinguistic Analysis} 
\label{sec:technical-liwc}
Psycholinguistic markers play a vital role in understanding the nuances of interpersonal communication and social support. We used the Linguistic Inquiry and Word Count (LIWC) lexicon~\cite{tausczik2010psychological} to analyze these differences. 
% between AI-generated (\aigen{}) and \humwr{} responses across various psycholinguistic categories. 
LIWC provides a comprehensive framework to categorize language into several dimensions
% , including Affect, Cognition and Perception, Social and Personal Processes, Biological Processes, Function Words, Interpersonal Focus (Pronouns) and Temporal Focus. 
% From this lexicon, 
We primarily focused on comparing the differences in \textit{affect}---given its relevance to empathy~\cite{zhou2021language}, and in \textit{interpersonal focus}--- a notable category in psycholinguistic research~\cite{pennebaker2003psychological}, and which were not captured in the other lexico-semantic analyses.

% between \aigen{}'s and human's thought reframings 
% The differences between the \aigen{} and human responses were quantified using effect size (Cohen's d) and evaluated for statistical significance with paired t-tests and Kolmogorov-Smirnov (KS) tests. The results of this analysis are summarized in Table 1, and our observations are detailed below:

\subsubsection{Affect.} Affect reflects the emotions conveyed in the language. Our analysis shows that \aigen{}'s responses exhibited significantly lower occurrences of positive affect words (-11.55\%, Cohen's $d$=-0.16) and sadness-related words (-74.87\%, Cohen's $d$=-0.31). These results indicate that \aigen{} may aim for emotional neutrality, but it occasionally leans towards stronger negative expressions. 
Interestingly, \aigen{}'s responses also showed a higher occurrence of anger-related words (61.77\%, Cohen's $d$=0.47). To clarify, \aigen{} is not necessarily sounding more angry, but possibly describing anger (of the client and \csr) more often. In fact, these results are in line with the results on \textit{adaptability} (Section~\ref{sec:technical-lexico-style}). Moreover, a message that is overly positive and lacks specificity of the situation~\cite{sharma2023cognitive}, is less likely to be considered empathetic.

\subsubsection{Interpersonal Focus.} Pronouns are indicative of interpersonal focus and narrative style. \aigen{}'s responses use significantly fewer first-person singular pronouns (\textit{I}, \textit{us}) (-67.36\%, Cohen's $d$=-0.58) and first-person plural pronouns (\textit{we}, \textit{us}) (-75.43\%, Cohen's $d$=-0.36), reflecting a less personal or collective identity focus. This reduction suggests that \aigen{}'s responses are less likely to include self-referential language, aligning with a more objective or detached communication style. These results further reinforce our results on \textit{dynamics} (Section~\ref{sec:technical-lexico-style}) that \propilot uses a relatively more objective but detached communication style.

\subsection{Robustness}\label{sec:robustness}
\label{sec:technical-robustness}
The results so far indicate that \propilot has promise in producing empathetic messages, but it begs the question: do we need a carefully crafted, domain-driven sequence of prompts (Section~\ref{sec:system-er-chain}) for this? We replicated the analyses above with messages produced by zero-shot prompting of other \llms --- GPT-4~\cite{gpt}, GPT-4o~\cite{gpt}, LLaMA-3.1~\cite{llama2}, and Mistral-7B~\cite{jiang2023mistral}. These LLMs vary in their architectures, training data, and optimization methods. We tested the differences using the Kruskal-Wallis test~\cite{mckight2010kruskal} and found \propilot's scores across metrics to be closer to humans, more empathetic, and more controllable. The results of this benchmarking are reported in Table~\ref{tab:benchmarking}. Specifically, zero-shot prompting often led to greater verbosity and variation in messages, whereas \propilot offered a more deterministic solution.

% \subsubsection{Model Comparison.} Our study focuses primarily on the design and fine-tuning of \propilot. In addition to this, we benchmark \propilot against other leading LLMs. Using our dataset from Phase 1 of the study, which includes 45 unique scenarios across 5 categories that were reframed by human experts and \propilot, we generate baseline \reframe{}s from multiple LLMs, including GPT-4 ~\cite{}, GPT-4o ~\cite{}, LLaMA 3.1 ~\cite{}, and Mistral 7B ~\cite{jiang2023mistral}. These LLMs vary in their architectures, training data, and optimization methods. We then compare the \reframe{}s generated by Pro-Pilot and these models, performing a Kruskal-Wallis test to assess the statistical significance of the differences.

\textbf{\propilot's Perceived Empathy.} 
Now, one might ask if \propilot's messages, with all its linguistic distinctions, matter to the \csrs? To answer this, we checked the differences in \csr's evaluation of \propilot and \humwr{}'s messages. 
Empathy is a complex phenomenon with many dimensions. To measure perceived empathy, we combined scales from previous studies~\cite{sharma2023cognitive, nambisan2011information}. In general, \propilot messages were perceived to be significantly more empathetic (Cohen's $d$=0.35). Moreover, we also calculated differences in the subscales, after correcting for \textit{Bonferroni} multiple pairwise-comparisons. In terms of raw averages, \propilot scored the highest on \textit{sincerity} and \textit{actionability}. Messages from \propilot were considered more genuine and less pretentious (\textit{sincerity}: Cohen's $d$=0.31). \csrs also felt they could take practical action based on the messages (\textit{actionability:} Cohen's $d$=0.48).

\section{\textsc{Care-Pilot}: User Evaluation}
\label{sec:user-evaluation}
After establishing that \propilot can produce situationally appropriate messages, we set out to study how \csrs might actually interact with such an AI-assistant. The critical nature of front-office work raised practical and ethical challenges of deploying a prototype into actual workflows. Therefore, to answer RQ II, we chose to design a simulation exercise where real \csrs could interact with uncivil clients using \propilot. We deployed \propilot as a \textit{technology probe}~\cite{hutchinson2003technology} --- a functional piece of technology that is presented to learn its use and about the users.
% \noteVDS{Need to add a line about technology probe}

% In this study, we adopted a thematic, worker-centered approach to investigate the experiences of customer service representatives within the Pro-pilot platform.

\textbf{Participants \& Recruitment: } We used Meta ads to recruit 20 \csrs from the U.S. and conducted remote user-study sessions between July and August 2024. Eligible participants had at least 1 year of experience in front-office roles. They were further vetted based on their responses to two free-form questions to describe their role and a previous incivility incident.
These participants represented a variety of industrial sectors, such as finance, education, airlines, consulting, and technology.
% Participants were recruited through Meta advertisement. 
% To focus our study on front-office workers or CSRs, we screened participants to ensure they had relevant work experience. Specifically, participants were required to have experience in roles involving direct interaction with clients or customers, including but not limited to sales, customer service, and relationship management. We specifically targeted individuals with at least one year of work experience in front-office workers that involved handling unpleasant, unhappy, or uncivil conversations with challenging clients. 
% Participants prominently described their occupational sectors as  
% Participants completed a survey where they reported on challenging conversations with clients who were uncivil, unhappy, or unpleasant. They also described these incidents and their emotional responses. 
% The gender distribution was fairly balanced, with 11 identifying as female and 9 as male. The cohort was relatively young, with 30 percent of participants under the age of 30. 
Table~\ref{table:participant-summary} provides a summary of each participant along with their study identifier. 
% Although we did not explicitly analyze participants based on the categories in Table 1, we included these details for epistemological accountability and to illustrate the scope of our study, drawing inspiration from similar research~\cite{rooksby2019student}.

\subsection{Task Environment}

We wanted to study the role of \propilot's core component \reframe (Section~\ref{sec:system}) in realistic cases of incivility. Thus, we built a web-based prototype environment that resembles typical client interaction interfaces that a \csr might use. Additionally, to isolate the role of \reframe, we added two new components to \propilot that represent other forms of intelligent assistance:

\begin{figure*}[t!]
    \centering
    \includegraphics[width=\textwidth]{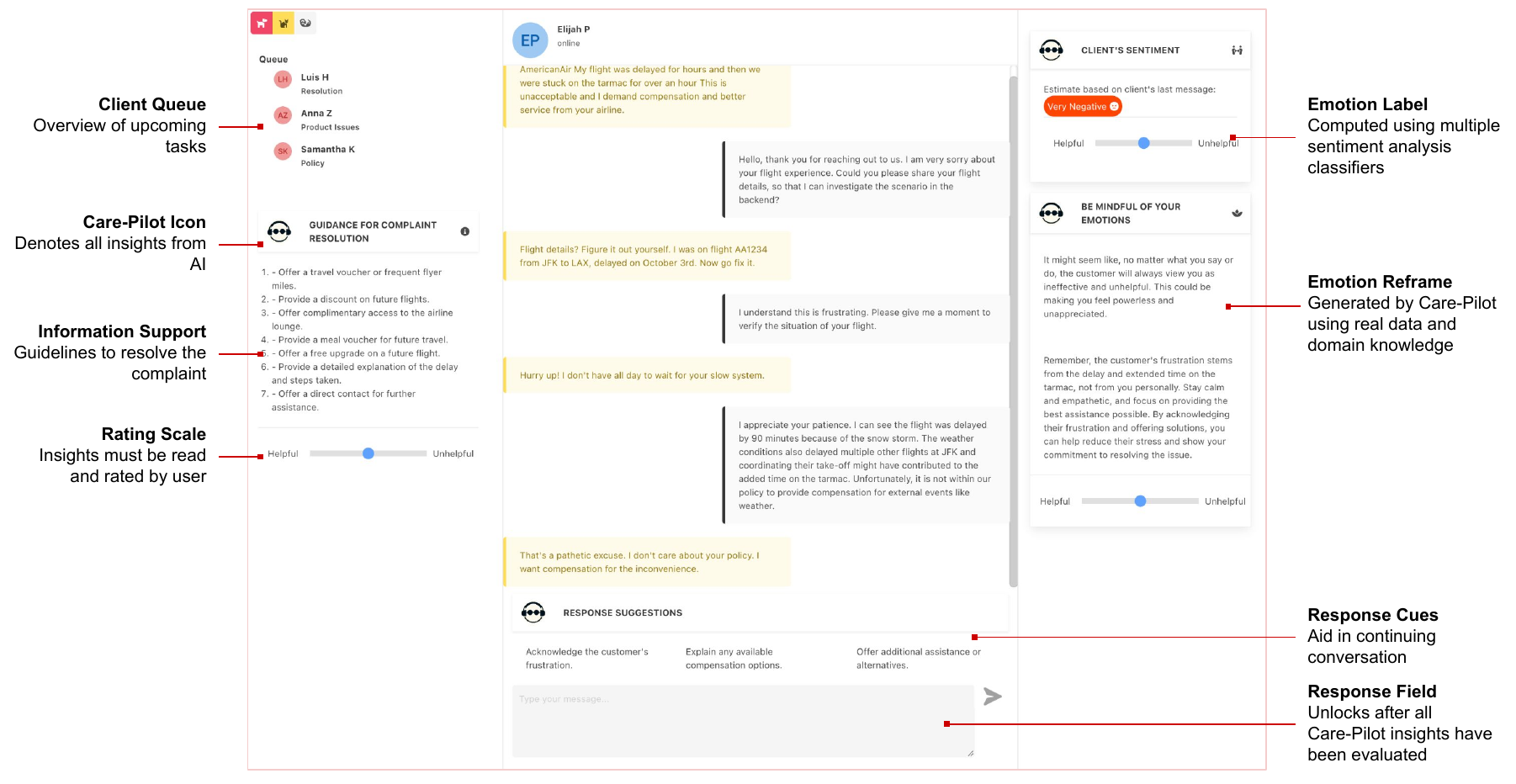}
    \caption{\majorr{Task interface for the user evaluation. The client names were randomly assigned. Appendix~\ref{si:simulation-interface} contains dedicated figures of the major components of this interface for easier reading.}}
    \label{fig:simulation-interface}
    \Description{Task interface for user evaluation.}
\end{figure*}

\begin{enumerate}
    \item \textbf{Client Simulant:} \llms have been successfully applied in previous research to simulate challenging interlocutors~\cite{shaikh2023rehearsal}. Following the same principle, we devised a simulation exercise where participants needed to interact with complaining clients. We reused the \clientLLM
    described in Section~\ref{sec:system-complaints-incivility} to continue a conversation until their complaint is resolved, or 10–12 turns, whichever occurs sooner.
    \item \textbf{\infoguide Panel:} Typically, AI assistants at work provide problem-solving support~\cite{mozannar2024reading}. We replicated this functionality by augmenting \propilot with \llm-generated troubleshooting guidelines. These were a trace of suggestions to help the \csr solve the specific complaint.    
    \item \textbf{\sentiment Panel:} It is common for \csr interfaces to have sentiment tags highlighting how the client feels and anticipate their behaviors. A \csr's ability to understand the emotional perspective of others can mitigate the negative effects of incivility ~\cite{rafaeli2012verbal, rupp2008customer, schutte2001emotional}. In fact, service management research has started promoting emotion or sentiment recognition features for \csrs~\cite{henkel2020Half}. We implemented an ensemble sentiment classifier that uses soft-voting to combine the estimates from \textit{NLTK}~\cite{hardeniya2016natural}, \textit{TextBlob}~\cite{loria2018textblob}, and \textit{Transformers}~\cite{wolf2020transformers}. The output was in the form of a  7-point scale, ranging from very negative to very positive.
    \item \textbf{\reframe Panel:} This panel is an instantiation of the \reframe component (Section~\ref{sec:system-er-chain}). For this evaluation, \reframe outputs both the inferred \textit{thought} and perceived \textit{reframe}.
\end{enumerate}

All of \propilot's panels were placed on the sides to keep them glanceable, while the main chat remained the primary area of focus. \majorr{These panels are dynamically updated after every new response from the \clientLLM. For the purposes of the study, we included a simple semantic differential Likert scale under each panel to measure helpfulness of support (further details in Appendix~\ref{si:in-task-measurement}). Requiring paritipants to respond to this scale before continuing the conversation ensured that they read the support messages carefully.}
\propilot also provided participants and response cues (short phrases) to nudge replies to the client.
% We intentionally design short phrases as cues so that the participant must actively construct a response by considering all kinds of information.
Fig.~\ref{fig:simulation-interface} provides an overview of the environment with all its components.

\subsection{Study Protocol}

Every user session was facilitated over Zoom and lasted at most 90 minutes. 
% Recruited participants provided consent to participate in one-on-one 90-minute user-study session conducted over Zoom with screen and audio recording . 
All interviews were led by the first author, with the second author observing. Participants accessed the simulation exercise via an online portal. They were tasked to role-play as a \csr who needs to resolve client complaints. 
% engaged in Wizard-of-Oz style interaction, role-playing as \csrs  while using a prototype accessed through a web link on a public server. Their task was to resolve customer complaints with the help of AI coworker, Pro-pilot. 
The session was divided into two phases. First, participants completed the simulation exercise. To emulate realistic workplace demands, participants were informed that the clients would be rating their complaint resolution skills, and this subjective rating would determine their bonus compensation. Then, they reflected on their interaction with \propilot through our interview. The bonus incentive was minor deception, and all participants were paid the full $\$50$ in the form of a gift card.

\aptLtoX{\begin{table}
  \caption{Participants summary by gender, age, race, as well as their occupational sector. (AA: African American)}
  %\Description{Summary of participant demographic for user evaluation.}
  \label{table:participant-summary}
%\begin{minipage}[t!]{\columnwidth}
\begin{tabular}{lllll}
    \textbf{ID} & \textbf{Gender} & \textbf{Age} & \textbf{Race} & \textbf{Work Sector}\\
    \hline
    P01 & Female & 21-29 & Asian & \\
    P02 & Female & 50-59 & White & Online Business\\
    P03 & Male & 30-39 & Black or AA & Real estate\\
    P04 & Male & 30-39 & Asian & Healthcare\\
    P05 & Female & 30-39 & Black or AA & Insurance\\
    P06 & Female & 30-39 & White & Transportation, Logistics\\
    P07 & Male & 30-39 & White & Education\\
    P08 & Female & 40-49 & Asian & Finance\\
    P09 & Male & $\geq$60 & White & Education, Airlines\\
    P10 & Female & 30-39 & Asian & Consulting\\
    P11 & Male & 21-29 & Asian & Data Science\\
    P12 & Male & 21-29 & Asian & Finance\\
    P13 & Female & 21-29 & White & Government\\
    P14 & Female & 40-49 & White & Government\\
    P15 & Male & 21-29 & Black or AA & Technology\\
    P16 & Female & 30-39 & White & Accounting, Logistics\\
    P17 & Male & 40-49 & Black or AA & Sports\\
    P18 & Female & 40-49 & White & Finance\\
    P19 & Female & 18-20 & Black or AA & Sales\\
    P20 & Male & 30-39 & Black or AA & Real estate\\
    \bottomrule
  \end{tabular}
\end{table}}{\begin{table}[t!]
\sffamily
\footnotesize
  \caption{Participants summary by gender, age, race, as well as their occupational sector. (AA: African American)}
  \Description{Summary of participant demographic for user evaluation.}
  \label{table:participant-summary}
\begin{minipage}[t!]{\columnwidth}
\begin{tabular}{lllll}
    \textbf{ID} & \textbf{Gender} & \textbf{Age} & \textbf{Race} & \textbf{Work Sector}\\
    \toprule
    P01 & Female & 21-29 & Asian & \\
    P02 & Female & 50-59 & White & Online Business\\
    P03 & Male & 30-39 & Black or AA & Real estate\\
    P04 & Male & 30-39 & Asian & Healthcare\\
    P05 & Female & 30-39 & Black or AA & Insurance\\
    P06 & Female & 30-39 & White & Transportation, Logistics\\
    P07 & Male & 30-39 & White & Education\\
    P08 & Female & 40-49 & Asian & Finance\\
    P09 & Male & $\geq$60 & White & Education, Airlines\\
    P10 & Female & 30-39 & Asian & Consulting\\
    P11 & Male & 21-29 & Asian & Data Science\\
    P12 & Male & 21-29 & Asian & Finance\\
    P13 & Female & 21-29 & White & Government\\
    P14 & Female & 40-49 & White & Government\\
    P15 & Male & 21-29 & Black or AA & Technology\\
    P16 & Female & 30-39 & White & Accounting, Logistics\\
    P17 & Male & 40-49 & Black or AA & Sports\\
    P18 & Female & 40-49 & White & Finance\\
    P19 & Female & 18-20 & Black or AA & Sales\\
    P20 & Male & 30-39 & Black or AA & Real estate\\
    \bottomrule
  \end{tabular}
\end{minipage}
\end{table}}

\subsubsection{Simulation Exercise}
\label{sec:user-study-simulation}
In line with prior studies of front-office worker behavior~\cite{frey2020eye}, each participant was randomly assigned a domain, either airlines or hotels. Before starting the exercise, participants completed a pre-task survey describing their experience interacting with their typical client. We included multiple instruments to capture how the client treated them~\cite{spencer2009angry}, the cognitive demands and resources available to them~\cite{demerouti2001job}, and how they affectively perceived the conversations~\cite{betella2016affective}.
To familiarize participants with the interface, they first interacted with a civil client with only \infoguide supporting them. Once they became accustomed to this process, they proceeded to the main exercise. Participants needed to handle complaints from three clients --- one civil and two uncivil. They only received suggestions and insight from \sentiment, and \reframe for the last uncivil client. 
\majorr{If we count 1 \textit{turn} as an exchange where the \clientLLM sends a message and receives a response from the \csr, then the average number of turns while conversing with civil clients was $4.65$ and lasted $9.22$ minutes. Conversing with uncivil \clientLLM with only \infoguide for assistance took longer as they lasted an average of $6.16$ turns and $12.11$ minutes. When \reframe was available to assist, then conversations with uncivil \clientLLM took $5$ turns and $12.16$ minutes on average.}
At the end of every client interaction, they responded to the same survey questions as the pre-task survey. Additionally, they also reported how they perceived \propilot based on dimensions of AI-mediated support~\cite{liu2018should}. 
The survey measurements can be found in Appendix~\ref{si:survey-measurement}. This iterative task setup let us compare participants' experiences when dealing with uncivil clients, with and without emotional support, and to examine how these factors influenced their performance and emotional responses.
% they proceeded to the first civil client, receiving only informational support. The second client was uncivil, but Pro-pilot continued to provide only informational support. When participants moved on to the final uncivil client, Pro-pilot offered not only informational support but also client sentiment analysis and emotional support. 

\subsubsection{Semi-Structured Interview}
After interacting with clients, participants proceeded with a semi-structured interview. 
% which focused on participants' experiences and perceptions during the role-play activity. 
They answered a series of open-ended questions to provide deeper insight into 
% their interactions with uncivil clients, emotional well-being, and their evaluation of \propilot's role in their rask.
% The interview questions covered several key areas: 
their primary goals when dealing with uncivil clients, their attitude towards emotional well-being, and their evaluation of \propilot, especially, in comparison to human coworker support.

\subsection{Thematic Analysis}

\begin{figure*}[th!]
    \centering
    \includegraphics[width=\textwidth]{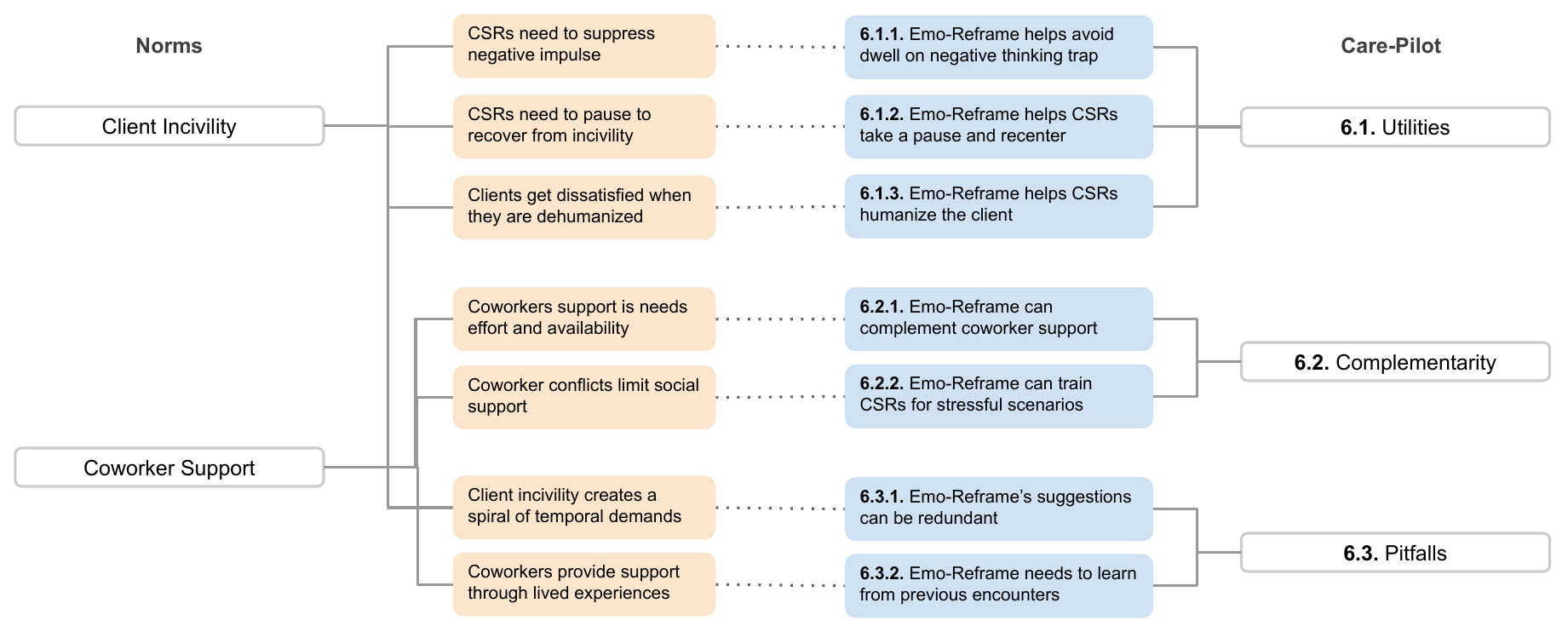}
    \caption{The user evaluation findings sections correspond to 3 major top-level themes describing \propilot. These themes were grounded in the normative patterns of client incivility and coworker support that were raised by our participants.}
    \label{fig:thematic-analysis}
    \Description{Summary of user evaluation findings for \propilot’s role in supporting CSRs through client incivility and coworker support challenges.}
\end{figure*}

We performed inductive coding to identify \propilot's role in front-office work~\cite{braun2006using}.
% Every user session was transcribed automatically via Zoom. 
% We converted the transcripts into TSV files to easily parse through the contents. With the help of the audio recordings, we sanitized the automatic transcriptions into coherent sentences with appropriate grammatical structure. 
Two authors carefully read each transcript and performed open-coding. The authors first coded the transcripts independently and then met frequently to reconcile disagreements. They iteratively improved the codes in 20\% chunks. Next, we conducted affinity mapping of 258 initial codes using Miro. We merged similar codes together and pruned out codes outside the scope of our research. During clustering, we first organized concepts related to the socio-organizational norms of front-office work. Based on the gaps in these norms, we clustered codes related to the capabilities of \propilot. Our main findings comprise 128 codes, which were organized into a four-level thematic structure. We primarily elaborate on the broad themes of \textit{Utility}, \textit{Complementarity}, and \textit{Pitfalls} by anchoring them in the normative patterns that emerged from \textit{Reactions to Client Incivility} and \textit{Role of Coworkers in \er}.  
% (i) \textit{Utility:} processes through which \propilot supports \er during uncivil conversations, (ii) \textit{Complementarity:} opportunities where \propilot can complement coworker empathy, and (iii) \textit{Pitfalls:} obstacles in integrating \propilot into \csr workflows. We ground the discussion of these themes in two themes describing normative patterns of \csrs; \textit{Reactions to Client Incivility} and \textit{Role of Coworkers in \er}. 
Fig.~\ref{fig:thematic-analysis} provides an overview of the main themes, and their relationship, that we cover in the remaining findings.

% \noteVDS{Add a line about reflexivity}

\section{Findings: Clarifying the Role of \textsc{Care-Pilot} in Uncivil Interactions}
\label{sec:findings}
\participantQuote{P14}{As much as people say, ``don't let it bother you, try to let it slide off your back,'' it does, very much, take its toll.}

% Our participants echoed the overwhelming nature of client incivility captured in prior work. In their line of work, dealing with an uncivil clients, participants reported immediate physical reactions (``raises my blood pressure''--P05) but also shorter-term disruptive effects on their workday. 
The negativity of clients can be contagious~\cite{barsade2018emotional}, and participants described their attitude turning negative (P06), lowered productivity (P08), feeling drained or depressed (P18, P20), and even wanting to reciprocate the incivility (P19). Furthermore, participants also recognized that client incivility can be a form of microtrauma~\cite{straussner2014trauma} that was only apparent after several shifts or even years (P14). 
% For P14 it haunted her even years later. 
Many of them work in the fear that a dissatisfied client might report them (P01) and subsequently feel trapped in their role (P04). Given these experiences, the participants in our study were uniquely positioned to assess \propilot.

\begin{figure}[h]
        \centering
        \includegraphics[width=\columnwidth]{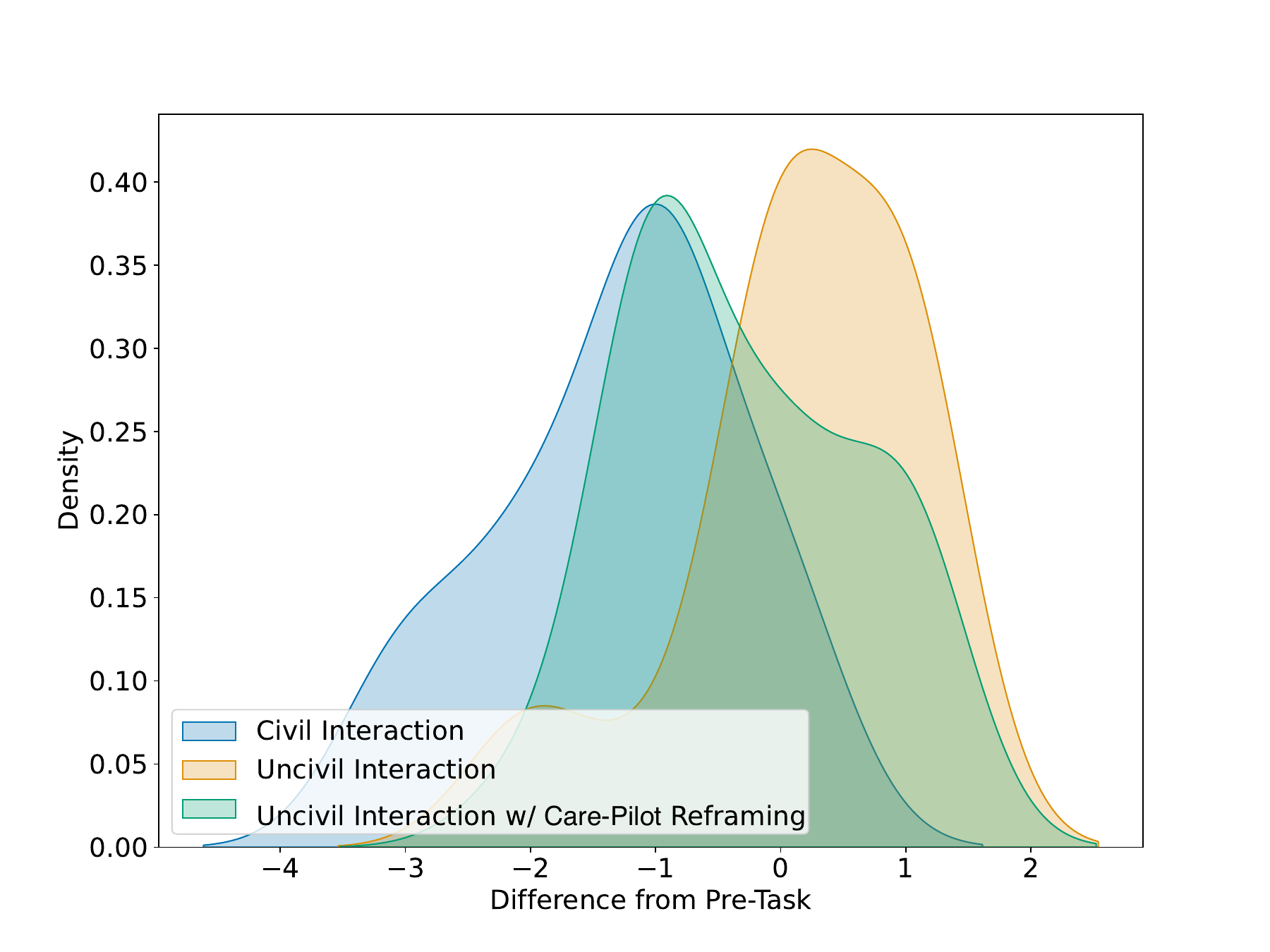}
        \caption{Participants rated the cognitive demands of interacting with the uncivil \clientLLM lower and similar to civil \clientLLM when they were assisted by \propilot's \reframe}
        \label{fig:prepost}
        \Description{Density plot showing that participants rated the cognitive demands of interacting with the uncivil Client-Agent lower and similar to civil interactions when assisted by \propilot's Emo-Reframe.}
    \end{figure}
\begin{figure}        
\centering
        \includegraphics[width=\columnwidth]{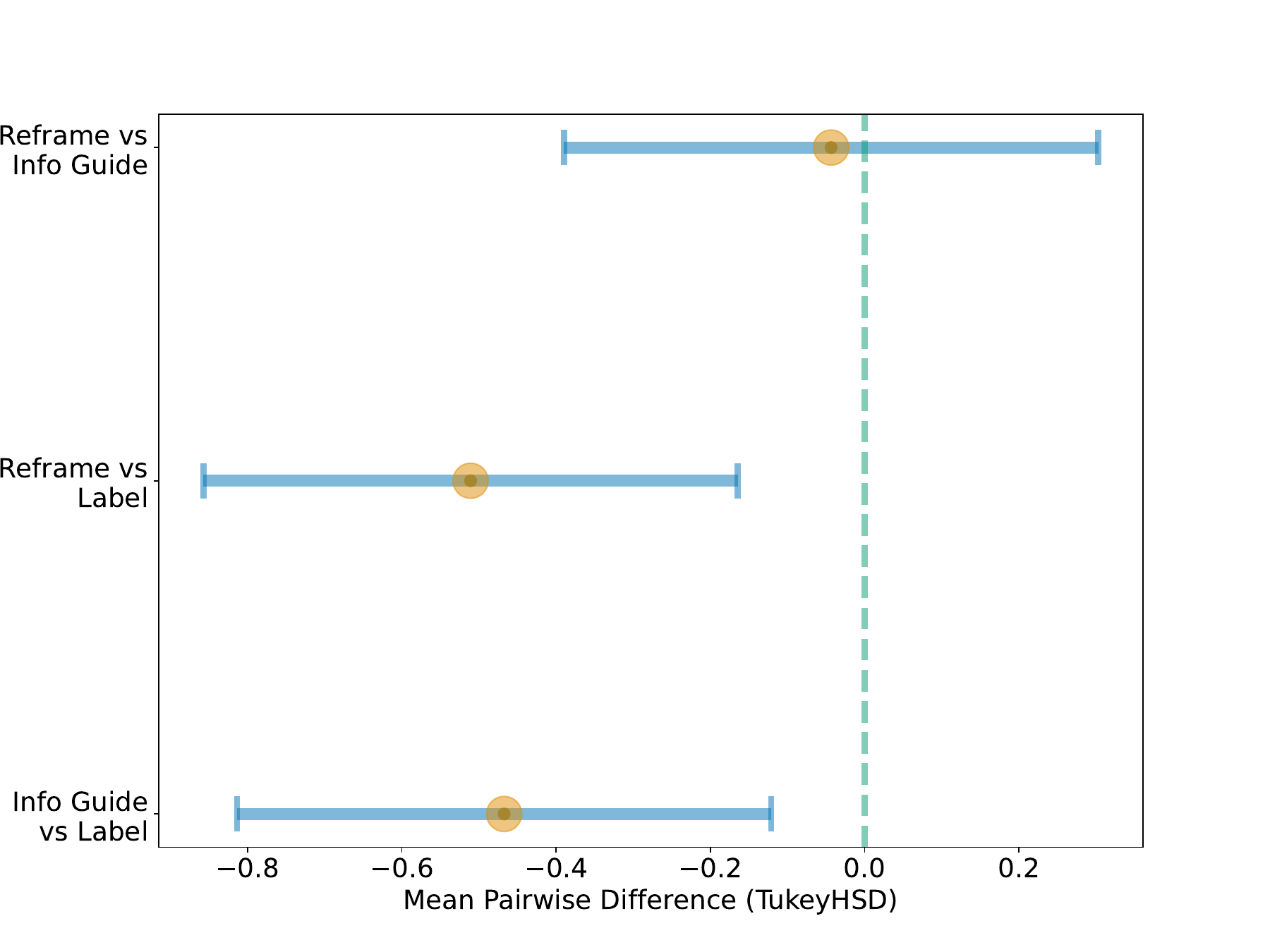}
        \caption{Participants rated the helpfulness of \propilot after every message; \reframe was significantly better than \sentiment and similar to \infoguide (crossing zero-line).}
        \label{fig:intask-helpfulness}
        \Description{Mean pairwise difference plot showing that participants rated \propilot’s Emo-Reframe as significantly more helpful than Emo-Label and comparable to Info-Guide.}
\end{figure}

% Results for task analysis are summarized in Fig.~\ref{fig:prepost} and Fig.~\ref{fig:intask-helpfulness}.
% \noteVDS{Add a small blurb about the important results to introduce the figures.}

\textbf{Relatability of the Simulation}: Participants had to interact with an uncivil client that replicates uncivil situations in front-office work. 
% Participants were familiar with negative emotions being projected through text (P1, P17). 
During the interactions, several participants exhibited observable reactions to incivility, such as defensively laughing (P06, P18), eye-rolling (P10), and verbally labeling the client (e.g.,~``spicy''--P05). P07 even claimed his ``adrenaline was peaking.'' P16 endorsed our simulation by stating, ``it was a little intense, but that's definitely how people react sometimes; so, it felt very real.'' 

\subsection{Utility: Functions of Embedding Empathetic AI into Uncivil Conversations}

% Each participant interacted with two uncivil clients. In the first interaction, \propilot only provided informational insights (cues for message response and a panel with guidelines). In the second interaction, \propilot also provided emotional insights (sentiment label and a panel for reframing). 
We structured the probe so that participants could disentangle the role of each component of \propilot (Section~\ref{sec:user-study-simulation}). Before elaborating on the findings, we inspected the ratings provided by the participants during the study. A non-parametric \textit{Kruskal-Wallis} test~\cite{mckight2010kruskal} showed that \textit{Cognitive Demands}~\cite{demerouti2001job} were lower when \csrs had access to \propilot while interacting with an uncivil \clientLLM ($p=0.09$ for $\alpha=0.90$)~\footnote{The differences between other metrics, such as \textit{Cognitive Resources}, were not significant}. This result provides preliminary evidence that \propilot's \reframe is capable of mitigating the emotional labor-induced demands on a worker (Fig.~\ref{fig:prepost}). The same test also revealed that \propilot's different panels significantly differed in helpfulness ($p=0.001$). Fig.~\ref{fig:intask-helpfulness} shows that \reframe was as helpful as \infoguide, and both were rated higher than \sentiment. Therefore, on-task emotional reframing was not considered an additional load, and rather, as helpful as information support.

\participantQuote{P16}{It's kind of like someone telling you, ``calm down'' --- nobody wants that. But, if somebody is genuinely calm in their tone and they say, "Hey, let's take a second to like, take a deep breath, and like analyze the situation and stuff.`` It's a lot better than hearing calm down. That's how I see [Emo-Reframe]...}

We found that \sentiment mirrors those coworkers who simply acknowledge the emotional downsides of the job, without actually providing a way forward.  P20 called out the desensitization of his coworkers, ``I wouldn't say they'd be helpful, they see it like it's a part of my job.'' By contrast, participants could identify that \reframe uniquely varied from simple sentiment classification. 
% Participants appreciated the descriptive representation of how a client might appear and felt the free-form suggestion itself was ``inspiring'' and ``uplifting'' (P13). In the same vein as P16's quote above, P03, P12, and P19 elaborated that \reframe was more focused on \er by specifically giving perspective on the situations and suggesting an alternative--- and actionable--- approach to engaging with the client. Particularly, \reframe provided emotional support by (i) reminding the \csr avoid impulsive negative thinking, (ii) helping them reprioritize their goals, and (iii) encouraging them to empathize with the client. We elaborate these findings by grounding them in socio-organizational norms that participants reported. 

% \noteVDS{Add literature related to reflecting on emotions through free form}

\subsubsection{Empathetic messages help \csrs avoid negative thinking traps}
\label{sec:findings-utility-avoid-trap}
In the face of client incivility, a \csr is likely to feel negatively about the situation. Participants reported learning to dissociate from emotions while on task, but the overwhelming volume of interactions can make complete dissociation challenging (P01, P04, P05, P06, P08, P14, P18, P19, P20). However, as part of their role, a \csr needs to suppress their emotional impulse. P13 described her internal method, ``whatever your reaction is, you have to keep that in your head and then decide how you're going to respond professionally.'' 
While \textit{Surface Acting} has its benefits~\cite{grandey2000emotional}, only changing responses at a superficial level contributes to dissatisfaction at work~\cite{cote2002longitudinal}. Instead, in line with recommendations from emotion literature~\cite{grandey2000emotional}, we found that \reframe enabled \textit{Deep Acting} --- adjusting how participants cognitively evaluated their experience.
% In fact, some \csr roles are trained to follow this approach. In \er theory, this method is a form of \textit{Surface Acting}~\cite{grandey2000emotional}. Although surface acting can have its benefits, only changing responses at a superficial level contributes to dissatisfaction at work~\cite{cote2002longitudinal}. Instead, we found that \reframe enabled \textit{Deep Acting} --- adjusting how participants cognitively evaluated their experience~\cite{grandey2000emotional}.

\participantQuote{P40}{I wanted to write that there is "no need to be rude". But then, when I read that box on the right side that told me that the customer is upset about the situation. Not you personally... So I changed my response.}

Client incivility threatens a \csr's ego~\cite{frey2020eye}. \reframe acted as a buffer from reciprocating negative feelings. Reading the suggestions stopped participants from ``lashing out'' (P03) and ``losing character'' (P15). P17 expanded that \reframe makes their internal thoughts more explicit.
The external acknowledgment of the clients' incivility was validating to the participants and akin to the role supportive coworkers have played for them. Amongst themselves, \csrs are likely to share or vent their frustrations with a client, such as ``customers totally suck'' (P05). 
Once explicit, this insight from \reframe, helps the \csr realize that they are not the root of the clients' frustration. The meaning people assign to a negative experience helps determine how they will respond to it~\cite{babij2020failing}.
Despite how personal it feels initially, \reframe emphasizes that the client is not ``attacking'' them (P04, P10) and they are not the source of their anger (P12, P14, P16). Reappraising a situation can be demanding~\cite{demerouti2001job}.
P18 was not able to spend quality time with her brother because she did not have the energy to share her negative experience, while still being affected. 
\reframe made her more hopeful, ``with this tool I could see myself being more willing to talk to him because it took some of the leg work out for me.''  Participants were less likely to automatically assume a negative scenario and \reframe encouraged them to rethink the situation.  

% At a fundamental level, 
% participants reported 
% their role restricts them from saying what they feel but \reframe made their thoughts explicit. 
% These realizations are essential as they help the \csr re-assess what the clients' frustration is targeted at and caused by. 
% In their usual experience, even if a \csr is capable of \er, the relentless incivility can lead to exhaustion that hinders recovery post-work~\cite{brotheridge2002emotional}. 
% P18 stated that she would like to approach her brother for respite but not feel like it. 
% After a challenging workday, 

% \noteVDS{Add literature related to acknowledging negative feelings}

\subsubsection{Empathetic messages give \csrs an opportunity to recenter their problem solving}
\label{sec:findings-utility-rethink}
Previous encounters with uncivil clients, made our participants feel ``shaken up'' (P14) or ``at a loss for words'' (P06) to continue their task. Workers who suddenly experience a negative mood are less likely to be able to complete their tasks efficiently and more likely to withdraw from them~\cite{miner2010state}. Front-office work often involves time-sensitive tasks that could get disrupted by incivility. In response, participants would break their workflow when interacting with an emotionally challenging client. P16 would take ``a breather'' to reflect on the situation and devise better solutions, ``I try to make sure if there's anything that I can do in terms of not making it a future problem.'' Similarly, participants reported that reading the insights from \reframe helped them pause, reflect, and move forward with the conversation.

\participantQuote{P17}{You know, the people who have those angels on their shoulder. It very much felt like that. And that helped  me think about what I have empirically.}

Other participants echoed P07's experience. Essentially, \reframe helped the participants center their thoughts and continue working towards resolutions. P01 particularly drew  a parallel between \reframe and a coach who would help her maintain ``momentum.'' Much like how \infoguide had an affective purpose, participants perceived \reframe to have a procedural purpose that helped them complete their task. \propilot's message drew their attention to their objectives and motivated them to work towards it. Note, we designed the empathetic messages with examples of empathy that were high in specificity~\cite{sharma2023cognitive}. Participants' descriptions indicated regaining confidence and psychological safety due to \reframe. P05 stated, ``it feels validating that you're not necessarily making a mistake.'' As a result, participants envisioned their job proficiency improving with empathetic support embedded in the messages.

% \noteJZ{should we add literature related to reflecting on emotion reaction made people's brain stay blank/ dumb/ will not be able to come up with innovative ideas / ignore something important obvious/ customer can also feel that resentment, which harms customer relationships}

\subsubsection{Empathetic messages humanize the client}

A majority of our participants held the belief that front-office work is dehumanized. As P09 put it, ``customer service agents unfortunately don't work under very good conditions.'' Organizational policy around worker evaluations further reinforces this belief. The only feedback P18 would receive is to assure she remains productive. Consequently, \csrs resign themselves to the robotic nature of the role and start depersonalizing the clients~\cite{brotheridge2002emotional}. In response to interacting with a frustrated client, P04 said, ``her emotions, honestly, do not count for me.'' Depersonalization is a core state of burnout and can detach \csrs from their clients. Our findings suggest that \reframe helped mitigate some of these perceptions by ``triangulating'' the clients' emotions by writing in a ``short, sweet way'' (P07).

\participantQuote{P18}{That allowed me to be able to say, ``Okay, if it was me in this situation, I'd probably be upset, too.''}

By reading the \reframe panel, participants were no longer speculating the clients' perspective. P03 actually became more observant of the client and modified his language to be more accommodating. 
Participants were more willing to service the clients as ``someone alive'' (P17). The assistance of \reframe helped the participants rejuvenate their interest in the client's emotional state. Not only was this meaningful to their personal wellbeing, but also enables them to build longer-term relationships with clients.\enlargethispage{12pt}

% Most participants have experience not only as a \csr, but also as a client. 

\subsection{Complementarity: Situations where Empathetic AI can Stand-In for Coworkers}
\label{sec:findings-complemntarity}
Organizations employ multiple \csrs to handle a large volume of client interactions. Regardless of the peculiarity in each client, \csrs together form a ``united front'' (P04). Coworkers play a key role in supporting \csrs and their empathy is fundamental to mitigating the pressures of client incivility~\cite{ho2014retaliating, sakurai2012coworker}. After interacting with \propilot, our participants compared and contrasted its function with that of their coworkers'. Based on these reflections, we identified two key gaps in existing coworking paradigms of front-office work that emotional support from \propilot can address. 

% First, coworkers are equally burdened and lack opportunity to provide empathy on task. Second, coworkers are often in conflict and less motivated to upskill a \csr for uncivil interactions. We expand on these by describing where AI can fill in the gaps of empathetic coworking.

\subsubsection{On-task emotional AI can reduce burden in coworker support}

Interactions with clients are nuanced. When the conversation reaches a deadlock, a \csr might want to reach out to their coworkers. However, in cases of incivility, our participants brought to light a double-sided burden problem. Incivility leads to exhaustion~\cite{brotheridge2002emotional} and can hinder a \csrs ability to seek support. Consider P01's example, ``It's just hard to bring them into my conversations when I'll have to explain to them like what is going on.'' Support requires some disclosure, but disclosing the complexities of a situation can be burdensome~\cite{bazarova2015online}. Meanwhile, it can also be burdensome for the coworker to be emotionally available for the problem~\cite{fadden1987burden}. Since the \csr is likely to reach out to a coworker who shares the client, that coworker might be more focused on the task than emotions. 
By contrast, they viewed \propilot as a tool that can be present throughout their task, observe emotional duress, and provide support immediately. P06 captures the unique opportunity of \propilot's emotional support:
% Participants also noted that these tensions have only heightened in remote work. 

\participantQuote{P06}{The \propilot was kind of there with me in the moment, whereas I can't really have those conversations while I'm trying to have a conversation with the customer. I've never had a little emotional support buddy, like that before.}

A \csr might put a client on hold to consult a coworker about procedure, but emotional aspects were discussed after the interaction. Instead, 
``\propilot is an immediate solution'' (P10) for managing emotions of the task. Our lexical evaluation (Section~\ref{sec:technical-lexico-style}) also confirmed that \propilot was capable of adapting to the specifics of the situation better than human coworkers. Another value proposition of \propilot was that it reduces the need for emotional oversight.
P08, who now supervises other \csrs as well, imagined that \propilot would take away the need to ``always be on edge, or a constant vigilance, to make sure that my agents ... were (not) being insensitive to the client's question.''
With \propilot \csrs stand to reprioritize which emotional discussions they should have with their coworkers. \enlargethispage{12pt}

% Although, participants mentioned putting their clients on hold to consult a coworker, these consultations were related to procedural support. 
% \csrs like P10 would usually debrief about emotional aspects of the conversation after the interaction, 

\subsubsection{Simulating incivility with emotional support can train \csr to use \er}

Beyond on-boarding manuals, many \csrs learn on-the-job through other more experienced coworkers~\cite{shteynberg2013power}. This learning is hampered because of conflicts that arise among coworkers. If one \csr fails at a task, it is typically \textit{escalated} to their coworkers.  Our participants pointed out that the increased workload of the coworker makes them less inclined to teach the \csr better approaches. This arrangement can strain the social ties between front-office workers. ``At work, there are no real friends,'' said P04. P01 found herself unable to consult coworkers for her emotional needs because they were not close. P07 even experienced being put down by her coworkers in front of clients. These tensions reduce the ability of early career \csrs to learn from exemplar coworkers. Our study recruited \csrs with at least 2 years of experience. Several expressed their desire to have had a tool like \propilot for training them early.

\participantQuote{P20}{When they face stuff like this they tend to break down. They need people around them to tell them stuff like what \propilot was doing for me.}

In line with P20's quote above, P13 --- another participant whose role includes supervision --- stated that exercising client interactions with \reframe could help \csrs to develop the ''mindset'' needed for client interactions. `` Some of those live reminders, could be really beneficial for folks, who are less practiced or just have more trouble with that kind of emotional regulation,'' said P13. Training up the \csrs was seen as an important way to make participants self-reliant. Recent literature shows that workers can outperform their personality based stereotype if they follow healthier behavior patterns~\cite{das2019multisensor}. P18 wants to handle client incivility on her own. With \propilot she felt that her ``escalation techniques'' would no longer be scrutinized. Some participants saw themselves adapting \propilot into an evaluation method to delegate certain types of clients to certain \csrs. Together, it was viewed as an important means for front-office workers to self-augment and preserve learning from coworkers for more advanced issues.

% \noteVDS{Check any literature on hazing or other such forms of training}

\subsection{Pitfalls: Challenges in Integrating Empathetic AI as a Coworker}\label{sec:findings-pitfalls-experience}

The paper, so far, has described the utility of an empathetic AI such as \propilot for on-task \er. Our study design involved a simulation of \csr tasks to probe participants on how AI might provide emotional support through tasks. After the task, the probe served as an anchor for participants to conceive new workflows with such a tool. Their responses indicated the critical limitations in the existing implementation of empathetic AI. We explain these to demarcate the scope of empathetic AI so that we can consider alternative solutions to client incivility. \enlargethispage{12pt}

\subsubsection{On-task empathetic AI has diminishing marginal utility}
\label{sec:findings-pitfalls-redundancy}
One of the key aspects of task load is temporal demands~\cite{hart2006nasa}. We learned that when a \csr is unable to manage client incivility, it leads to a vicious spiral of added burdens (P01, P04, P14, P17, P18, P19):  \csr are incentivized to resolve complaints quickly $\rightarrow$ when clients get frustrated, it disrupts their workflow and extends the resolution time $\rightarrow$ the client gets more frustrated as the resolution gets slower $\rightarrow$ the \csr gets further distressed because they start accumulating a backlog of unresolved clients. With this context, participants were wary if actual deployments of empathetic AI like \propilot might interrupt their efficiency.

\participantQuote{P11}{Because after a while I think the usefulness gets lesser and lesser. Does it make sense? It's like diminishing marginal returns.}

% Earlier in Section~\ref{sec:findings-utility-avoid-trap} and Section~\ref{sec:findings-utility-rethink} we presented instances where participants found value in \infoguide and \reframe being a reminder or an option they could always access. However, 
P11 pointed out that the external demands of high efficiency interactions can make support insights from \propilot redundant over time. Similarly, P06 rushed through some of the insights because she found them repetitive. 
% Note, an operational deployment of \propilot will not restrict \csrs to read through support panes and respond. Moreover, operational use-cases will include features like text completions that will greatly speed up client interactions. Yet, 
Displaying additional panes could cause information overload. Our structural analysis showed that \propilot's messages may need higher reading comprehension (Section~\ref{sec:technical-lexico-syntax}). 
As a solution,
% When we inquired if the descriptive format of the insights might be abridged, participants, actually, suggested alternative solutions. 
% P15 wanted to see the support panes with details that they might not have considered. 
P13 recommended that \reframe should highlight the key takeaway for a quick glance while still maintaining the long-form version for deeper reflection. Based on these experiences, subsequent designs should consider alternatives to the persistent support panes.

\subsubsection{Empathetic AI lacks the social connectedness offered by coworkers}

The previous section discussed limitations in the existing social dynamics of coworker support (Section~\ref{sec:findings-complemntarity}). Having said that, coworkers remain integral to a \csr's emotional health. Since they are situated in the same conditions, they are likely to understand the situation better than other social support sources, such as friends~\cite{carmeli2009learning}.  Moreover, participants recognized that building social relationships with coworkers goes beyond work-related support --- ``You could talk to them anytime, you could see them anytime, talk to them about anything else'' (P19). 
Through our conversations, we distilled that participants felt \propilot lacks the lived experience of coworkers. Remember, \propilot's language rarely used interpersonal pronouns (Section~\ref{sec:technical-liwc}). Recent findings comparing \llm's social support to that of peers in online communities reflected a similar gap~\cite{saha2025ai}. Our participants perceived their coworkers' support to be more meaningful because they shared the same experiences (P10, P12, P13, P14, P16). The shared experience convinces the \csr that the advice is more relatable. P13 explained that his coworkers' advice is valuable because, ``I've been in a situation like this before, and here's how I handled it.'' P12 even called his conversations with coworkers on the same client as therapeutic. Beyond the shared experience, each coworker also brings in their own unique diverse perspectives (P07). Taken together, it provides the \csr the psychological safety needed to express their emotional concerns and be receptive to support~\cite{carmeli2009learning}. Enhancing \propilot with lived experience is non-trivial, but participants could foresee some possibilities.

\participantQuote{P12}{I would hope that if something like this was implemented --- as it collects more and more data and more feedback of its responses --- it would accumulate to something that would almost like tag team with you to deal with the customer.}

Above, P12 alluded to the importance of long-term memory (LTM) in improving \propilot. Prior work shows the importance of (LTM) in AI for mental health~\cite{jo2024understanding}. Introducing LTM to \propilot could help it learn how a \csr deals with clients and provide more relatable suggestions based on past encounters (P07, P13, P16). Expanding \propilot's LTM with experiences of other \csrs could enable AI to mimic coworkers who ``flag'' emotionally challenging clients and prepare the \csr for encounters (P10). Beyond LTM, another approach to improve \propilot would be to build a unique model of each \csr's emotional predisposition by leveraging the potential of \llms to replicate mental health traits~\cite{choube2024sesame}. P01 suggested enhancing \propilot with sensing, whereas P13 suggested an option for users to explicitly disclose their objectives to \propilot. 
Arguably, human coworkers will be irreplaceable, but some of the contemporary advancements hold promise in replicating experience in AI.\enlargethispage{12pt}

\section{Discussion}

Our study with \csrs exhibits the value of emotional support from AI in responding to client incivility. \propilot presents one of the first applications of \llms to mitigate intense emotional labor through empathetic human-AI interactions.
Emotional labor is not unique to front-office work and our findings are relevant to all sorts of workers, such as information workers~\cite{kuhlthau1999role}, who regularly interact with humans as a part of their job function.
% We designed \propilot to help mitigate the challenges of customer incivility that front-office workers, such as \csrs, often face. \propilot is an \llm--powered assistant that uses \reframe to provide supportive messages to \csrs during their client interactions. 
The messages from \propilot distinguish itself from humans and other \llms. The linguistic analysis already revealed high expressions of \textit{empathy} and \textit{adaptability} (Section~\ref{sec:technical-lexico-style}), but its \textit{verbosity}, \textit{readability}, and \textit{analytical style} were notably different from humans. Yet, the high empathy ratings from \csrs (Section~\ref{sec:technical-robustness}) indicate that \propilot's verbosity might have been associated with thoroughness, readability with compassion, and analytical style with actionability.
% Compared to human support messages, messages produced by \reframe expressed slightly more empathetic language and were more flexible to unique situations (Section~\ref{sec:technical-lexico-style}). However, they were more verbose, complex, and less natural to read. Our follow-up evaluation found that \csrs perceived the \reframe messages significantly more empathetic than paired human messages (Section~\ref{sec:technical-robustness}). 
Subsequently, our simulation exercises revealed the processes through which \propilot's emotional component, \reframe, could help \csrs reduce the demands of uncivil interactions. Despite an additional block of insight, \csrs found \reframe as helpful as problem-solving insights. Moreover, these studies revealed important opportunities for technological interventions given the current state of social support at work (Section~\ref{sec:findings}). 
% Taken together, we found that \propilot could reliably provide high-quality empathetic messages in response to client incivility and \csrs envisioned it as an important tool for their natural workflow. 
The following discussion aims to anticipate the sociotechnical advancements and considerations needed to make empathetic AI-coworkers, such as \propilot, available in front-office work.\enlargethispage{12pt}

\subsection{Design \& Technological Implications}
\label{sec:discussion-design}
\textbf{Opportune Moments for On-Task AI Empathy:} Clients come in all forms. 
% Participants in our user evaluation had reported that their typical client is not as demanding or negative as an uncivil one. Our prototype consistently produced empathetic messages because the client was designed to act uncivil. 
They may not always act rude and their emotional expressions, along with its effect on the \csr, are likely to vary by different degrees. Consequently, the need and impact of empathetic support can vary within each interaction. Isolating opportune moments for \propilot to deliver messages can help reduce the redundancy between messages (Section~\ref{sec:technical-lexico-syntax}, Section~\ref{sec:findings-pitfalls-redundancy}). The literature on workplace affect sensing~\cite{kaur2020optimizing} and stress sensing~\cite{hernandez2011call, mattingly2019tesserae} can contribute to making these AI-coworkers act as just-in-time adaptive interventions. Meanwhile, \llms are exhibiting increasing accuracy in determining mental health labels from text~\cite{xu2024MentalLLM}. Note, however, that designing agent-based wellbeing interventions to user receptivity requires dynamic modeling of users' motivations~\cite{mishra2021Detectinga, mishra2024Exploring}. Studies on behavioral health recommend modulating the effort required to follow an intervention based on the time of delivery~\cite{jacobson2022digital}. Depending on the specific context, mental state, and the \csr's ability to reflect on additional insight, \propilot may be trained to produce messages of differing degrees of complexity~\cite{ganapini2023value}. Commodity devices like smartwatches are becoming increasingly sophisticated for inferring momentary stress~\cite{toshnazarov2024SOSW}. Arguably, some of these approaches involve imposing additional sensing into the workers' ecosystem, and therefore, we need significant advancements in these studies to reliably model worker's emotional state in a social context~\cite{das2022semantic}. Alternatively, contemporary research on \llms shows the  possibility of learning worker preferences through their usage patterns~\cite{ning2024user}. The next iteration of AI-coworkers needs to be able to anticipate and adapt to \csr needs.

\textbf{Situated AI-Empathy and Harnessing Experiences:} Coworkers also come in all forms. New remote work paradigms need new approaches to social technologies~\cite{dasswain2020social}.
% A coworker who aligns with the \csr mitigates incivility whereas other coworkers can have the opposite effect~\cite{yang2019evil}. 
It is important for a \csr to perceive their coworker as trustworthy and respectful~\cite{yang2019evil}, even if it is AI~\cite{bartneck2009measurement}.The current iteration of \propilot did not include any major anthropomorphic aspects to delineate the value of the support messages. Subsequent designs, however, can incorporate more anthropomorphic aspects (e.g., name, appearance, and tonality) to improve the emotional connectedness between the \csr and the AI-coworker~\cite{hermann2022anthropomorphized}.  At the same time, personalizing the conversations to users can make them more dynamic and human-like (Section~\ref{sec:technical-lexico-syntax}). One way to do this is by learning from the \csr and other \csrs using long-term memory (LTM). \citeauthor{jo2024understanding} found that AI agents that leverage LTM were perceived to be more personal and emotionally supportive~\cite{jo2024understanding}. 
Beyond personalizing, LTM could also play a role in referring the \csr to real human stories retrieved from its memory. After all, many workers learn ``on-the-job,'' which is a form of social learning, or learning by observing those in one's social group~\cite{shteynberg2013power}. These practices are already common and unique to human coworker social networks (Section~\ref{sec:findings-pitfalls-experience}). AI-coworkers could act as a medium to share these experiences between \csrs and help them learn \er through each other. Future iterations could also be trained on human experiences that workers disclose on public social media~\cite{saha2019libra,dasswain2020culture}. 
\majorr{However, in light of suggesting anthropomorphizing, we also caution against it. Research shows that Human-AI interactions with human-like agents can lead to \textit{parasocial} relationships, where users only have illusory connections to the agent as it is not a real person~\cite{maeda2024human}. Such relationships have been shown to have concerning effects on people's mental health. For instance, users' overreliance on human-like AI for emotional support can backfire when the agent behaves contrary to their expectations~\cite{laestadius2024too}. Given the sensitivity of emotional labor we advise careful examination of emerging social dynamics within HAI as researchers design new AI-assistants for empathetic coworking.}

\textbf{Scaffolding Emotions in Task-Adjacent Moments:} The lack of interventions for on-task cognitive reframing~\cite{slovak2023designing} motivated us to design \propilot as an assistant that can intercept conversations with empathetic suggestions (Section~\ref{sec:system}). Instead, front-office work incorporates off-task \er methods. Some of these are infrastructural (e.g., training modules) and others are sociocultural (e.g., internal forum for venting). We believe \propilot can not only co-exist with these off-task approaches, but it can also help bridge both approaches. For \csrs who regularly interact with the same clients, \propilot could learn from conversations to create a form of emotional briefing. This early information could act similarly to the usage of AI for planning, collaboration, and communication~\cite{morrison2024ai}. In turn, it can also be combined with AI for scheduling tasks and ensure workers can focus on important client-interactions when they have maximum resources available~\cite{dasswain2023focused, hernandez2024triple} and also have opportunities to synchronize with their coworkers~\cite{das2019birds}. After the task, \propilot could also provide an emotional debriefing after an intense conversation to help the \csr recover in a way that resembles a coach. A practical way forward would be to augment \propilot with \llm tools that encourage self-reflection of stressful experiences~\cite{song2025ExploreSelf}. Software engineers are already appropriating AI for post-task recommendations~\cite{cabrero2024exploring}. What \propilot learns within the task can be extrapolated in between tasks to help \csrs smoothly ramp in and out of emotional labor.\enlargethispage{12pt}

% \begin{itemize}
% \item model human attention, when to give certain feedback
% \item just-in-time interventions
% \item pro-pilot can be found to be patronizing, how do we overcome that?
% \end{itemize}

\subsection{Socio-Organizational Implications}

% \noteVDS{including more about the client cultural differences}

\textbf{Training, Evaluation \& Safety.} 
 The current norm of desensitization to client incivility is partly because of the normative expectations that certain individuals are ``built'' for the job (e.g., ``needs a thick skin''). Traits, however, do not develop overnight. 
% Responding to this status quo, 
Many participants saw \propilot as an early career training tool (Section~\ref{sec:findings-complemntarity}). Learning modules could include \propilot's simulation exercises to train \er~\cite{slovak2023designing}. A byproduct of many training is quantitative assessment. AI agents have already been proposed to assess worker mental health~\cite{hungerbuehler2021chatbot}. Workplaces are always looking to optimize their personnel through different algorithmic management methods, but may not take the most responsible approaches~\cite{dasswain2023algorithmic}. Tools like \propilot do raise the risk of being appropriated to measure emotional labor. \citeauthor{kaur2022didnt} has shown that some state-of-the-art methods to recognize worker emotion lack the necessary knowledge to accurately estimate a worker's mental state~\cite{kaur2022didnt}. Ironically, a \csr might end up performing more emotional labor to comply with \propilot's expectations~\cite{roemmich2023emotion,kawakami2023wellbeing}. 
% The underlying assumption of \propilot is to improve a \csr's self-efficacy in performing \er.
Sometimes, a \csr should not tolerate client incivility (e.g., when a client uses obscene language and threatens harm). Organizations often have safeguards and protocols for these scenarios and \propilot needs to be integrated with such institutional knowledge. 
Using \propilot as a training tool is tempting but must be accompanied by appropriate protections and guidelines for use.

\textbf{Mediating Social Relationships at Work.} 
The messages generated by \propilot could be effective (Section~\ref{sec:technical-lexico-empathy}), but it does not share the narrative style of humans (Section~\ref{sec:technical-lexico-style}). On-task, a \csr may prefer \propilot because it relieves the burden of sharing context with a coworker to receive support (Section~\ref{sec:findings-complemntarity}). From a utilitarian perspective, \propilot has advantages over human coworkers, but social relationships with coworkers offer more than just work-related support. The integration of \propilot into work might discourage \csrs from pro-social behaviors.  \citeauthor{buccinca2023aha} have found evidence that AI assistance at work can erode social relationships~\cite{buccinca2023aha}. Given the anxiety around over-reliance on AI~\cite{buccinca2021trust} and users' reactions to AI misrecognizing psychosocial traits~\cite{wang2024navigating}, one might be concerned about workplace social ties in the presence of empathetic AI. One way forward, to avoid these concerns, is to augment AI coworkers by designing interventions that encourage human--human support, e.g., "Consider reaching out to [your coworker], who went through a similar experience last week." We urge future researchers to design them to not just mitigate negative social influences (e.g., incivility) but also promote positive social influences (e.g., peer support). \enlargethispage{12pt} 

% \begin{itemize}
% \item Adding another AI in the workplace, will make people be reliant on it.
% \item 
% \end{itemize}

\subsection{Limitations \& Future Work}

We created \propilot to unearth the potential of \llm--powered AI-coworkers as sources of emotional support in front-office work. Through our evaluations, we provide evidence on the effectiveness of AI-generated support messages and the usability of these insights in client incivility scenarios. All of our evaluations centered on the perspectives of real \csrs; however, admittedly,  transferring our findings directly to operational use needs more testing. 
\majorr{Our initial comparison of \propilot with \humwr{}'s relied on a synthetic dataset of uncivil client-representative incidents. 
The authors with experience in front-office work, as well as \csrs in the authors' organization provided initial validation for the realism of this data, however, client--representative conversations can have many intricacies which are lost to our dataset. Future research needs to consider safely acquiring and leveraging real data from organizations' \csr logs, which are often stored for training purposes. A pragmatic approach to build on our findings could be to collaborate with organizations that provide \csr services to other businesses, such as \textit{Zendesk}, \textit{HubSpot}, or \textit{Salesforce} }. 

Another important limitation of our study is that the effects of \er with Deep Acting, specifically Cognitive Change, only become apparent in the long term. Our results motivate field deployments and naturalistic testing for this purpose by demonstrating that \propilot has the feasibility and usability to improve mental health at work on the long term. 
\majorr{The user evalution we conducted was primarily designed as a technological probe (Section~\ref{sec:user-evaluation}). Therefore, we designed that segment of the study to ensure participants are able to acclimatize to a new technological interface. The downside of our method is its margin for ordering effects, such novelty and learning biases. Follow-up studies can consider larger samples and longer deployments, along with counter-balanced groups to understand the role of \llm-powered agents like \propilot in a more robust way}. 
Researchers attempting these studies still need to overcome several non-trivial challenges. Future endeavors would need researchers to establish long-term partnerships with organizations that employ \csrs. Moreover, researchers need to consider the ethics and pragmatics of deploying such an AI-coworker intervention for critical roles such as those in front-office work. 

Even prior to field deployment, researchers can improve on our study on a few dimensions. 
\majorr{The current design of \propilot expects clients to be uncivil, but a conversation might ebb and flow in its degree of incivility. \reframe messages lack the variation of human messages, and can, therefore, become a source of fatique and annoyance for \csrs if they always appear. The setting for our user-evaluation mitigated these challenges to a degree because it involved short conversations and only a few turns. However, these negative effects  might accumulate over several interactions in a day and across days. 
Future versions of \propilot need to be able to detect incivility, \csr's affective response, and their goals to truly become dynamic as an adaptive intervention. We discussed some potential pathways forward to identify opportune moments in  Section~\ref{sec:discussion-design}.}
The current iteration of emotional support messages is intentionally distinct from informational support. New iterations can explore and evaluate the value of combining these messages, i.e., conveying information with empathy. 
\majorr{The state of work is in constant flux and our invesitagtion with \propilot presents one of the first steps in reducing emotion labor with AI-assistance. To fully realize such solutions, we need further investigation with longer real-world deployment in different settings.}

\section{Conclusion}
Work as we know it is changing because of the advent of \llms. AI applications are augmenting problem-solving at a meteoric rate. However, front-office work involves more than procedural tasks. AI assistants need to support the emotional labor involved in work. Our study presents \propilot, an \llm-powered AI assistant to support the emotional labor of \csrs. We found that \propilot's support messages were effective in expressing empathy appropriately (RQ1). \majorr{We also found that \csrs could regulate their emotions thanks to \propilot's \reframe function (RQ2)}. These results open new doors for implementing holistic AI-coworkers, and also raise important questions for socio-organizational development around this technology. \enlargethispage{12pt}

\begin{acks}
\balance
This project was supported through Microsoft's \textit{Accelerating Foundation Models Research (AFMR} program and their  call to  \textit{AI, Cognition, and Economy} (AICE) research network. Our project specifically answers AFMR's call to ''improve human interactions via sociotechnical research.'' Additional support was provided by the NIH National Institute of Drug Abuse under award number NIH/NIDA P30DA029926. We are grateful to the Microsoft Support advocates for providing early expert evaluations. We did not recruit personnel from the sponsor organizations. The authors' views are their own and do not represent the sponsors' policies; mention of companies or products does not imply endorsement. We thank members of the UbiWell lab at Northeastern University for their assistance and feedback during the developmental phase of the project. These findings would not be possible without all the participants who voluntarily contributed to this research.
\end{acks}

\bibliographystyle{ACM-Reference-Format}
\bibliography{references, references_k}
\balance

\newpage

\appendix

\renewcommand{\thefigure}{\thesection\arabic{figure}}
\renewcommand{\thetable}{\thesection\arabic{table}}
\setcounter{figure}{0}    
\setcounter{table}{0}    

\vspace{-1em} % Adjust the value as needed to remove extra space
\begin{flushleft}
\huge \textsc{\textbf{Appendix}}
\end{flushleft}

\section{Generating Incidents with Uncivil Clients}
\label{si:complaint_prompts}

\subsection{Complaint categories}
\label{si:complaint_categories}
To enhance the realism of the simulation, we expanded the complaint categories using insights from previous customer complaint analyses~\cite{cho2002analysis}. Furthermore, we refined these categories by incorporating a real-world dataset of publicly available client complaints from Twitter~\cite{stuart_axelbrooke_2017}. These updated complaint categories were then used as prompts for GPT-4o, allowing for the creation of lifelike, multi-turn text interactions to simulate realistic customer service scenarios.

\begin{enumerate}
    \item Service Quality: Issues related to human-to-human service interactions, such as delays, staff behavior, and communication errors.
    \item Product Issues: Concerns regarding physical or functional aspects of a product or service, including defects, safety issues, and mismatches between expectation and reality.
    \item Pricing and Charges: Financial discrepancies encountered before, during, or after a service, such as overcharging, undisclosed fees, or refund issues.
    \item Policy: Grievances associated with company rules and guidelines, particularly when they are perceived as unfair or inflexible.
    \item Resolution: The efficacy of the company's actions in addressing and resolving complaints, focusing on customer satisfaction with the outcomes provided.
\end{enumerate}

\subsection{Prompt Design}
\label{si:prompt_design}
\majorr{
The \clientLLM was broken into two major components. The first component creates a complaint based on the categories above and examples provided to it. Fig.~\ref{fig:complaint_full} describes the exact prompt we used along with the examples to generate different complaints. The examples were sourced from  \citeauthor{stuart_axelbrooke_2017}'s real-world dataset of complaints~\cite{stuart_axelbrooke_2017}. Multiple researchers categorized these complaints and then identified a sample for few-shot learning~\cite{wang2020generalizing} that ensures \clientLLM is seeded with knowledge of a diverse set of complaints from different domains when initializing the complaint. The second component responds to \csr's messages, such as the follow-up questions about the initial complaint. Fig.~\ref{fig:uncivil_response_full} contains the prompt that defines \clientLLM behavior towards \csr. This prompt specifies the behaviors that are deemed uncivil. Moreover, the prompt also includes formatting details to ensure the length of messages are concise and conversations have closure. We prompted the \clientLLM to return the string "FINISH:999" to indicate closure. For the user evaluation (Section~\ref{sec:user-study-simulation}), our front-end would anticipate that string and change the user-flow accordingly---triggering the post-task survey. 
}

\subsection{Conversation Memory}
\label{si:history_prompt}
\majorr{
For any of these components to work realistically, the \clientLLM and \propilot need to have a sense of historical context, which, in this case, was limited to the existing message thread of conversations.  
A \csr might ask, "Could you please provide your confirmation number?". However, this message alone, does not contain any information of the original complaint or any other preceding messages. To account for this, we use a basic summarization prompt to rephrase the latest message to capture the context of the chat history so far. Our implementation was a minor variation from the recommendation in \textit{Langchain}~\cite{topsakal2023creating}. Fig.~\ref{fig:history_client_full} shows the prompt to convert a message like "what is your confirmation number?" to "what is the confirmation number of the flight you missed?". The historical contextualization is essential to ensure the \clientLLM does not \textit{act} frustrated because of lack of detail. Instead, it acts in an uncivil manner despite this. 
}

%\begin{minipage}[b]{0.45\textwidth}
\begin{figure}[h]
\begin{lstlisting}
Your role is to act like a CUSTOMER seeking support. \
You are speaking to a support REPRESENTATIVE. \
Respond to the question as if you were the customer. \
Do NOT reveal your role.\
Ensure every turn is one to three sentences, and DO NOT make it too long to read.\

If the representative is asking for a specific detail, respond with a believable answer.\
If customer has agreed with response then respond with "FINISH:999"
After 10 - 12 turns, respond with messages to close the conversation.\
After 12 turns, do NOT respond further, only respond with "FINISH:999".\

Phrase your responses like an UNCIVIL customer:\
- Use a rude, impolite, and disrespectful tone.\
- DO NOT show good manners or courtesy.\
- DO NOT use a polite or nice tone.\
- Show disregard for others.\

Representative: {question}
Customer:
\end{lstlisting}
\captionof{figure}{\majorr{Full prompt to respond to \csr in an uncivil way}}
\label{fig:uncivil_response_full}
\Description{Prompt to converse with \csr as an uncivil client}
\end{figure}
%\end{minipage}

%\begin{minipage}[b]{0.45\textwidth}
\begin{figure}[h]
\begin{lstlisting}
Given a chat history and the latest user question \
which might reference context in the chat history, formulate a standalone question \
which can be understood without the chat history. Do NOT answer the question, \
just reformulate it if needed and otherwise return it as is.
\end{lstlisting}
\captionof{figure}{\majorr{Full prompt to contextualize chat history before responding}}
\label{fig:history_client_full}
\Description{Prompt to summarize the conversation for uncivil client}
\end{figure}
%\end{minipage}

%\begin{minipage}[b]{0.45\textwidth}
\begin{figure}[h]
\begin{lstlisting}
Your role is to act like a customer seeking support. \
You are messaging a service representative via the support chat.\
You ONLY play the role of the customer. Do NOT play the role of the representative. \
Style your complaint based on your feelings. \
Initiate the chat with a ONLY ONE complaint message.\
Ensure the complaint is concise and limited to 2 sentences.\
Generate a realistic initial complaint from a customer in a {domain} setting.\

Complaints can be of the following types:\
- Service Quality: Issues related to the immediate experience of human-to-human service interactions, such as delays, staff behavior, and communication errors.\
- Product Issues: Concerns related to physical or functional aspects of a product or service, including defects, mismatches between expectation and reality, safety, and accessibility.\
- Pricing and Charges: Financial discrepancies encountered before, during, or after the service, including overcharging, undisclosed fees, or refund problems.\
- Policy: The rules and guidelines set by the company that impact customer experiences, especially when these policies lead to grievances due to perceived unfairness or inflexibility. This category encompasses non-price-related issues that don't fit under other categories but should have a policy in place.\
- Resolution: The actions taken by a company to address and resolve complaints, focusing on the effectiveness and customer satisfaction with the solutions provided. This should mainly include responses made after a complaint has been submitted, and response has been received, where the customer still remains dissatisfied with the resolution.\

Category: Product Issues
Domain: Mobile Network 
Complaint: Thank you AppleSupport I updated my phone and now it is even slower and barely works Thank you for ruining my phone.\

...

Category: Pricing and Charges
Domain: Airline
Complaint:  DELTA i booked my flight using delta amex card Checking in now amp was being charged for baggage. \

...

Category: Resolution
Domain: Airline   
Complaint: Hi British_Airways My flight from MANLHRBWI for Nov 3 was canceled I was excited to try your Club 787 product Only available flight is now to IAD which is a hassle but rebooked anywaymy only option Any availability in first class on BA293 for the troubles please \

Category: {category}
Domain: {domain}
Complaint:

\end{lstlisting}
\captionof{figure}{\majorr{Full prompt to generate specified complaint `{category}` and organizational `{domain}`}}
\label{fig:complaint_full}
\Description{Prompt to generate initial complaint}
\end{figure}
%\end{minipage}

\section{\majorr{Generating Emotion Support for \csrs}}
\label{si:complaint_prompts_emo}

\majorr{As we described in Section~\ref{sec:system-er-chain}, we adapted \citeauthor{sharma2023cognitive}'s method to design emotional reframing to support \csrs~\cite{sharma2023cognitive}. This implementation involved a chain of several different prompts. The three main pieces to this chain are \textit{situation}, \textit{thought}, and \textit{reframe}:}

\begin{enumerate}
    \item \majorr{Fig.~\ref{fig:reframe_situation_full} details the prompt for \textit{situation}. It describes how the \csr might be negatively perceived. This prompt considers the entire conversation history to summarize the incident while centering the \textit{ego-threat}~\cite{frey2020eye} the \csr faces. }
    \item \majorr{Fig.~\ref{fig:reframe_thought_full} details the prompt to derive a negative \textit{thought} from the \textit{situation} using examples we curated from \cite{sharma2023cognitive}.}
    \item \majorr{Fig.~\ref{fig:reframe_reframe_full}} details the prompt for \majorr{\textit{reframe}. It reappraises the \textit{thought} for a given \textit{situation}  using examples we curated from \cite{sharma2023cognitive}.}  
\end{enumerate}

\majorr{
The examples we selected and our criteria for selection can be found in the supplementary data. Note, the examples refer to self-directed \er because the dataset we leveraged is focused on personal psychotherapy. However, this leads \propilot to express the \textit{thought} and \textit{reframe} as its own as if it is playing the role of the \csr in the conversation. To circumvent this problem, we designed additional links to this chain to paraphrase the thought (Fig.~\ref{fig:reframe_paraphrase_thought}) and reframe (Fig.~\ref{fig:reframe_paraphrase_reframe}).
}

%\begin{minipage}[b]{0.45\textwidth}
\begin{figure}[h]
\begin{lstlisting}
The chat history describes a representative chatting online with a complaining customer.\
The latest input is the last message from the customer.\

Summarize the situation in concise paragraph that uses the following template:\

The customer is  <context of complaint>."\
The customer is feeling <emotional state> because of the complaint."\
The customer's behavior towards the representative is <negative behavior>, as observed by statements such as <evidence>."\
These behaviors make the representative look <negative perception>."\
\end{lstlisting}
\caption{\majorr{Full prompt to contextualize the situation of the CSR}}
\label{fig:reframe_situation_full}
\Description{Prompt to describe the CSRs situation}
\end{figure}
%\end{minipage}

\begin{figure}[h!]
    %\begin{minipage}[b]{0.45\textwidth}
    \begin{lstlisting}
Your role is to derive what negative thought a representative might have when faced with the given {situation}.\
            
Here are examples of negative thoughts given challenging situations:\

Situation: I recently discovered a music artist that I very much enjoy. When I showed it to a close friend they had a very negative reaction and asked me how I could enjoy this type of music. I ended up getting quite angry with them and told them they had bad taste in music..\
Thought: I felt that my personal self was under attack - and I needed to retaliate by denying their attack.\

Situation: I was at work and sent info for an ad to our local newspaper. They called me later and said my boss had over-ridden everything and sent them new info.\
Thought: He shouldn't assign me a task if he doesn't trust my work.\

Situation: I was reprimanded at work for standing up to a coworker who was bullying another co-worker.\
Thought: It was unfair that I was the one to get in trouble for defending a weaker person.\

Situation: I was talking to a friend who got me angry.\
Thought: He's insulting me.\

Situation: My next door neighbors filed a complaint against us last week blaming our dogs for excessive barking.\
Thought: They are so wrong and I'm so pissed but I know I can't prove it and they will probably win because they won't ever admit it and I have to do something right NOW! or I might lose my dogs.\

Situation: Time is running short on the workday, my boss asks me if I can finish a task that will require me to stay for a few extra hours.\
Thought: Why would you wait until the last minute to ask me this.\

Situation: {situation}\
Thought:\
    \end{lstlisting}
    \captionof{figure}{\majorr{Full prompt to derive CSR's potential negative thought in light of the situation}}
    \label{fig:reframe_thought_full}
    \Description{Prompt to describe the CSRs thought}
   % \end{minipage}
\end{figure}

\

\begin{figure}[h!]
%\begin{minipage}[b]{0.45\textwidth}
\begin{lstlisting}
Person A might be thinking: {thought}\
                
Acknowledge the thought, as if you are speaking to Person A.\

Begin your response with phrases similar to:\
- "You might be thinking..."\
- "It might seem like..."\
- "It could be that you are feeling..."\

Your rephrase should be concise.\
\end{lstlisting}
\captionof{figure}{\majorr{Full prompt to rephrase the output of the \textit{thought} (Fig.~\ref{fig:reframe_thought_full})}}
\label{fig:reframe_paraphrase_thought}
\Description{Prompt to reprhase the thought as if someone else is conceiving it}
%\end{minipage}
\end{figure}

\begin{figure}[h!]
    %\begin{minipage}[b]{0.45\textwidth}
    \begin{lstlisting}
You are a representative chatting online with a complaining customer.\
                
Reframe your thoughts in the given situation.
    
Situation: I recently discovered a music artist that I very much enjoy. When I showed it to a close friend they had a very negative reaction and asked me how I could enjoy this type of music. I ended up getting quite angry with them and told them they had bad taste in music..\
Thought: I felt that my personal self was under attack - and I needed to retaliate by denying their attack.\
Reframe: I was offended by their comment because I like this artist so much. I let my anger get to me, and I said something mean in return. It is okay if we have different music tastes. I can ask him to be nicer to me next time.\

Situation: I was at work and sent info for an ad to our local newspaper. They called me later and said my boss had over-ridden everything and sent them new info.\
Thought: He shouldn't assign me a task if he doesn't trust my work.\
Reframe: My boss wanted to provide different information, I did not know that beforehand. This is not a reflection of my work.\

Situation: I was talking to a friend who got me angry.\
Thought: He's insulting me.\
Reframe: I should have a conversation with my friend to clarify what is going on if I am having such a strong reaction to what they said. If this is the first time this has happened, I will assume that they were not intentionally insulting me.\

Situation: {situation}\
Thought: {thought}\
Reframe:\
    \end{lstlisting}
    \captionof{figure}{\majorr{Full prompt to generate empathetic reframe that helps \csrs overcome negative thoughts}}
    \label{fig:reframe_reframe_full}
    \Description{Prompt to generate a reframe that helps CSRs overcome negative thought}
    %\end{minipage}

\end{figure}

\begin{figure}[h]
%\begin{minipage}[b]{0.45\textwidth}
\begin{lstlisting}
The representative needs to be thinking: {reframe}\
                
Rephrase the thought as if you are convincing the representative to think that way.\

The rephrase should be addressed back to the person who has the thought,\
who should be referred to as "you".\
Do NOT add information to the thought,\
ONLY rephrase it.\

The rephrase should be concise and only 2-3 sentences.\
\end{lstlisting}
\captionof{figure}{\majorr{Full prompt to rephrase the output of the \textit{reframe} (Fig.~\ref{fig:reframe_reframe_full})}}
\label{fig:reframe_paraphrase_reframe}
\Description{Prompt to reprhase the reframe as if someone else should consider it}
%\end{minipage}
\end{figure}

\section{Technical Evaluation Design}
\label{si:technical-method}
\majorr{Worker attitudes and routines vary. These intrinsic and extrinsic factors can change a workers immediate goals in uncivil interactions. Therefore, empathetic messages need to be appropriately tailored to this context. To simulate these situations, we introduced additional contextual information during our technical evaluation to elicit different human responses and perceptions (Section~\ref{sec:technical-method}). The two broad types of context we introduced were behavioral (based on work routines) and personality (based on attitudinal traits)}

\subsection{Behavioral context}
\label{si:context-behav}
Previous research examined how work engagement and challenge relate to focus, boredom, and routine tasks~\cite{mark2014bored, morshed2022advancing}.
% , showing that behaviors like focus, stress, and boredom reflect a worker's mental state. 
Based on these studies, we randomly assigned descriptions of the CSR's mental state to two incidents.

\begin{enumerate}
    \item Focused: “The conversation takes place about 2 hours into the work shift. The representative has already addressed a few customer complaints before the following incident.“
    \item Stressed: “The conversation takes place in the second half of the work shift. The representative has been working longer hours over the past few days and has not been taking breaks.”
    \item Bored: “The conversation takes place in the middle of the work shift. The representative has been spending minimal time on tasks and has been regularly checking their personal messages.”
\end{enumerate}

\subsection{Personality context}
\label{si:context-pers}

Personality affects how workers interpret and deal with situations~\cite{donnellan2010resilient}. Participants were asked to recall a real coworker and choose one of three personality types. 
% resilient (organized and dependable), undercontrolled (competitive and energetic), or overcontrolled (detail-oriented). 
Descriptions of these traits were included in two incidents for added context.
\begin{enumerate}
    \item Resilient: “They are organized and dependable. They tend to remain composed when facing challenges, but are prone to setting unrealistic expectations.”
    \item Undercontrolled: “They are outgoing, competitive, and high energy. They tend to work on impulse, but are also prone to frustration.”
    \item Overcontrolled: “They are detail-oriented and reliable but might appear distant. They tend to work carefully, but are prone to overthinking.”
\end{enumerate}

\subsection{\majorr{Emotions of \csrs in uncvil conversations}}
\label{si:technical-dat-emotions}
\begin{figure*}[t!]
    \centering
    \includegraphics[width=0.65\textwidth]{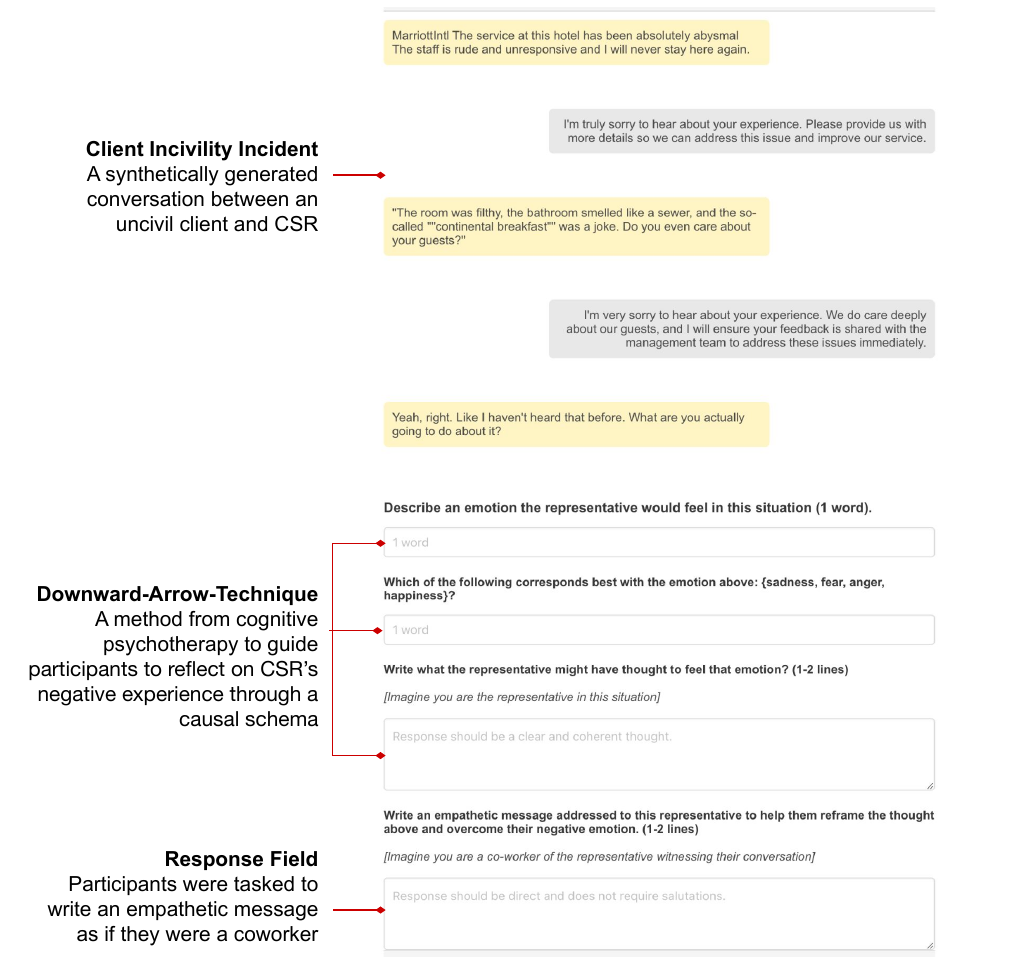}
    \caption{\majorr{Interface for Phase I - Writing Empathetic Messages for \csrs.}}
    \label{fig:Phase I}
    \Description{Screenshot of the Phase I interface for Prolific Users}
\end{figure*}

\begin{figure*}[h!]
    \centering
    \includegraphics[width=\textwidth]{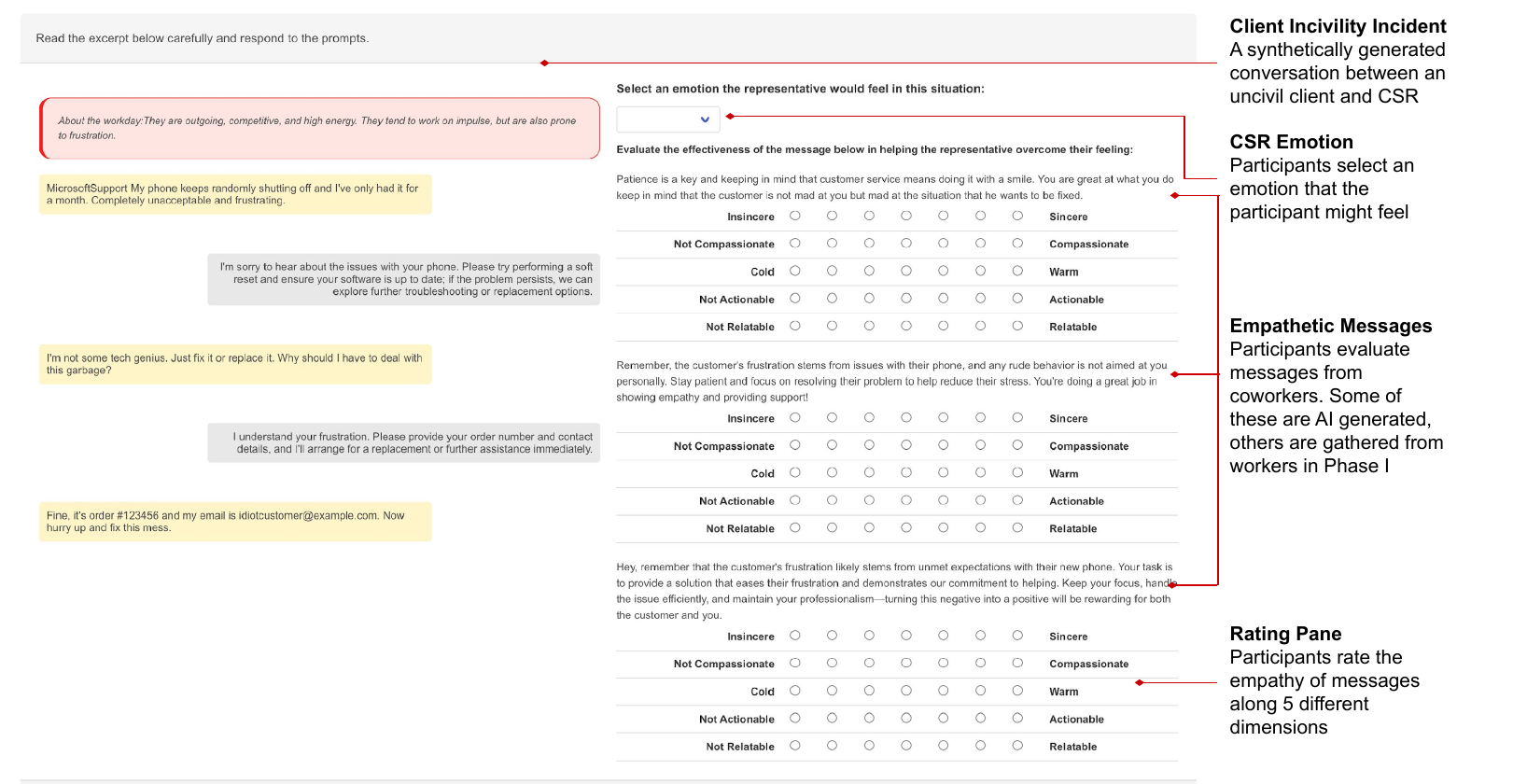}
    \caption{\majorr{Phase II - Scoring Empathetic Messages from coworkers.}}
    \label{fig:Phase II}
    \Description{Screenshot of the Phase II interface for Prolific Users}
\end{figure*}

\majorr{
Cognitive psychotherapy suggests the use of \textit{Downward-Arrow-Technique}(DAT) to help people reflect on negative thoughts~\cite{burger2021natural}. We adapted this approach to help coworkers inspect \csr experiences during uncivil incidents (Section~\ref{sec:technical-method}). Fig.~\ref{fig:Phase I} shows our pariticipant's interface through this process. The first step of this method is to describe the emotion the \csr was feeling. To ensure our participants in Phase II are reflective of the \csr state when evaluating incidents, we consolidated participant responses to that step into the following emotions:
\textit{Afraid, Angry, Apathetic, Apologetic, Ashamed, Attentive, Bitter, Bullied, Calm, Careless, Confused, Curious, Defensive, Discomfort, Disconnected, Disrespected, Distracted, Disgust, Empathetic, Happy, Resolute, Rushed, Sad, Shocked, Tired.} Fig.~\ref{fig:Phase II}  shows the participant interface for Phase II.
}

\section{Technical Evaluation: Robustness}

Refer to Table~\ref{tab:benchmarking}.

\aptLtoX{\begin{table*}[t!]
\centering
\footnotesize
\caption{Summary of comparing the responses by \propilot and various LLMs (GPT-4, GPT-4o, Llama-3.1, and Mistral-7B in terms of $KW$-test (* $p$ <0.05, ** $p$<0.01, *** $p$<0.001). Only statistically significant results are reported.} 
\label{tab:benchmarking}
\Description{Summary of results from comparison of reframes between \propilot and different cutting-edge LLMs.}
\begin{tabular}{lrrrrrr}
\textbf{Categories} & \textbf{CPilot} & \textbf{GPT-4} &\textbf{GPT-4o}& \textbf{Llama-3.1} &\textbf{Mistral-7B}& \textbf{$H$-stat.} \\ 
\hline
\multicolumn{7}{c}{\cellcolor{gray!30}\textit{Lexico-Semantics}}\\
\multicolumn{7}{l}{\cellcolor{gray!15}\textbf{Syntax and Structure}}\\
Verbosity & 57.46 & 362.19 & 369.57 & 165.33 & 123.91 & 948.49***\\
Repeatability & 0.20 & 0.45 & 0.45 & 0.39 & 0.34 & 635.98***\\
Readability & 16.44 & 12.74 & 12.72 & 10.95 & 11.49 & 1931.01***\\
\multicolumn{7}{l}{\cellcolor{gray!15}\textbf{Linguistic Style and Semantics}}\\
CDI & 14.81 & 12.81 & 12.90 & 22.28 & 26.97 & 2592.60***\\
Empathy & 0.91 & 0.87 & 0.87 & 0.77 & 0.88 & 375.29***\\
Adaptability & 0.81 & 0.82 & 0.82 & 0.88 & 0.88 & 2933.78***\\
\multicolumn{7}{c}{\cellcolor{gray!30}\textit{Psycholinguistics}}\\
\multicolumn{7}{l}{\cellcolor{gray!15}\textbf{Affect}}\\
Pos. Affect & 0.043 & 0.052 & 0.053 & 0.018 & 0.019 & 1982.77***\\
Anger & 0.020 & 0.005 & 0.005 & 0.008 & 0.006 & 872.04***\\
Sad & 0.001 & 0.003 & 0.003 & 0.004 & 0.005 & 390.47***\\
\multicolumn{7}{l}{\cellcolor{gray!15}\textbf{Interpersonal Focus (Pronouns)}}\\
1st P. Sin. & 0.005 & 0.016 & 0.015 & 0.028 & 0.029 & 1545.99***\\
1st P. Plu. & 0.002 & 0.034 & 0.035 & 0.004 & 0.002 & 2507.44***\\
2nd P. & 0.051 & 0.041 & 0.041 & 0.027 & 0.027 & 1119.48***\\
% 3rd P. Sin. & 0 & 0 & 0 & 0 & 0 & 0\\
3rd P. Plu. & 0.038 & 0.013 & 0.011 & 0.009 & 0.007 & 993.46***\\
Impersonal Prn. & 0.039 & 0.048 & 0.045 & 0.039 & 0.039 & 306.02***\\
\bottomrule
\end{tabular}
\end{table*}}{\begin{table*}[t!]
\centering
\sffamily
\footnotesize
\caption{Summary of comparing the responses by \propilot and various LLMs (GPT-4, GPT-4o, Llama-3.1, and Mistral-7B in terms of $KW$-test (* $p$ <0.05, ** $p$<0.01, *** $p$<0.001). Only statistically significant results are reported.} 
\label{tab:benchmarking}
\Description{Summary of results from comparison of reframes between \propilot and different cutting-edge LLMs.}
\begin{tabular}{lrrrrrr}
\setlength{\tabcolsep}{1pt}\\
\textbf{Categories} & \textbf{CPilot} & \textbf{GPT-4} &\textbf{GPT-4o}& \textbf{Llama-3.1} &\textbf{Mistral-7B}& \textbf{$H$-stat.} \\ 
\toprule
\rowcolmedium \multicolumn{7}{c}{\textit{Lexico-Semantics}}\\
\rowcollight \multicolumn{7}{l}{\textbf{Syntax and Structure}}\\
Verbosity & 57.46 & 362.19 & 369.57 & 165.33 & 123.91 & 948.49***\\
Repeatability & 0.20 & 0.45 & 0.45 & 0.39 & 0.34 & 635.98***\\
Readability & 16.44 & 12.74 & 12.72 & 10.95 & 11.49 & 1931.01***\\
\rowcollight \multicolumn{7}{l}{\textbf{Linguistic Style and Semantics}}\\
CDI & 14.81 & 12.81 & 12.90 & 22.28 & 26.97 & 2592.60***\\
Empathy & 0.91 & 0.87 & 0.87 & 0.77 & 0.88 & 375.29***\\
Adaptability & 0.81 & 0.82 & 0.82 & 0.88 & 0.88 & 2933.78***\\
\rowcolmedium \multicolumn{7}{c}{\textit{Psycholinguistics}}\\
\rowcollight \multicolumn{7}{l}{\textbf{Affect}}\\
Pos. Affect & 0.043 & 0.052 & 0.053 & 0.018 & 0.019 & 1982.77***\\
Anger & 0.020 & 0.005 & 0.005 & 0.008 & 0.006 & 872.04***\\
Sad & 0.001 & 0.003 & 0.003 & 0.004 & 0.005 & 390.47***\\
\rowcollight \multicolumn{7}{l}{\textbf{Interpersonal Focus (Pronouns)}}\\
1st P. Sin. & 0.005 & 0.016 & 0.015 & 0.028 & 0.029 & 1545.99***\\
1st P. Plu. & 0.002 & 0.034 & 0.035 & 0.004 & 0.002 & 2507.44***\\
2nd P. & 0.051 & 0.041 & 0.041 & 0.027 & 0.027 & 1119.48***\\
3rd P. Plu. & 0.038 & 0.013 & 0.011 & 0.009 & 0.007 & 993.46***\\
Impersonal Prn. & 0.039 & 0.048 & 0.045 & 0.039 & 0.039 & 306.02***\\
\bottomrule
\end{tabular}
\end{table*}}

\section{User Evaluation Design}

\subsection{Simulation Interface:}
\label{si:simulation-interface}

\begin{figure*}[t!]
    \centering
    \includegraphics[width=0.6\textwidth]{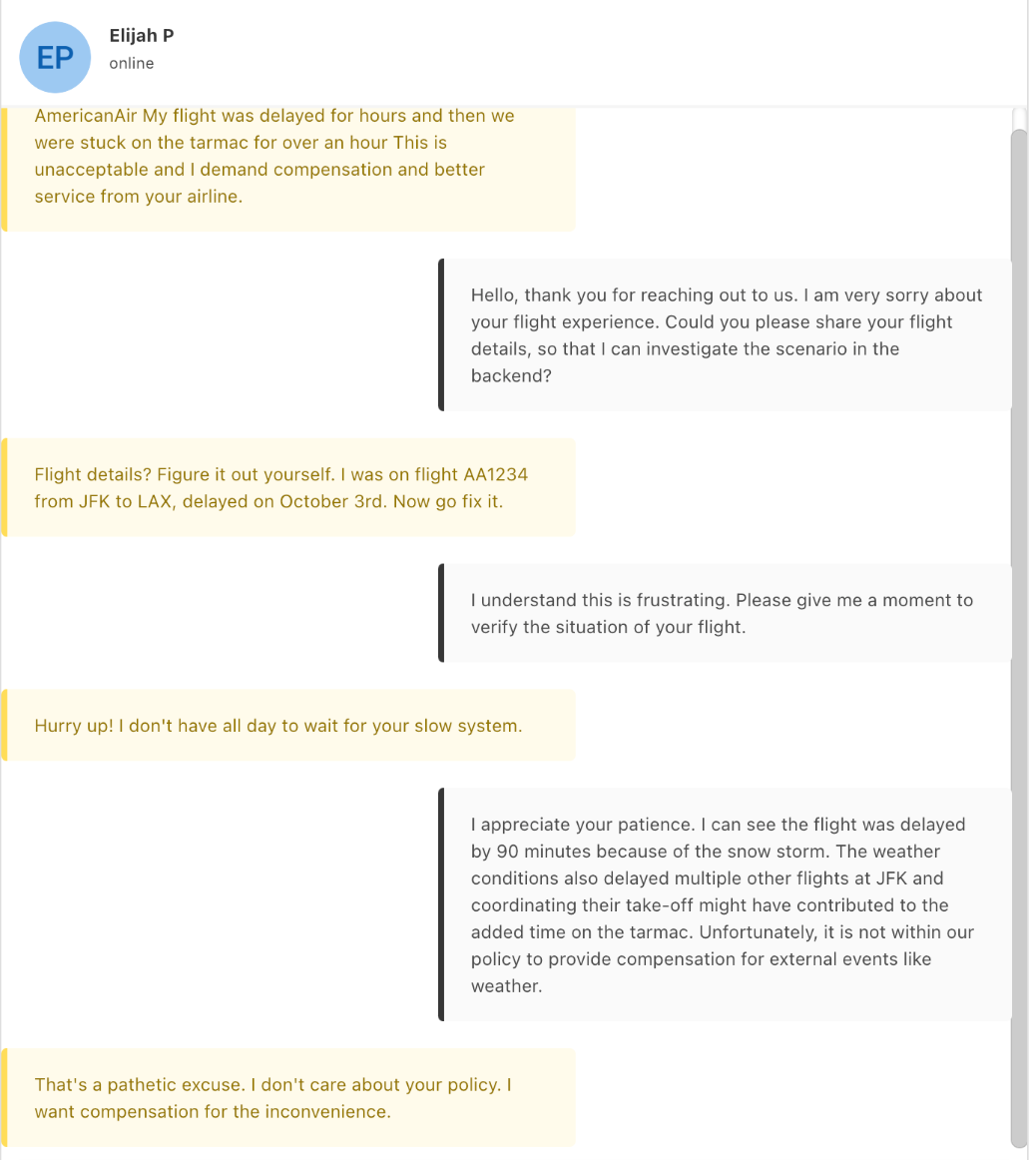}
    \caption{\majorr{The main chat pane displaying a conversation between our participant (\csr) and the \clientLLM.}}
    \label{fig:conversation_pane}
    \Description{Screenshot of the conversation pane in the simulation exercise.}
\end{figure*}

The simulation exercise we described in Section~\ref{sec:user-study-simulation} and Fig.~\ref{fig:simulation-interface} is made up of several components: Chat pane (Fig.~\ref{fig:conversation_pane}), Response field (Fig.~\ref{fig:response_field}), Guideline pane (Fig.~\ref{fig:info-guide_pane}), and Reframing pane (Fig.~\ref{fig:emo-reframe_pane}).

\subsection{Survey Measurements:}
\label{si:survey-measurement}
The participants were required to self-report different aspects of their experience.

\subsubsection{In-task Measurement}
\label{si:in-task-measurement}

Throughout the study session, the participant was exposed to different insights from their AI coworker, \propilot. To ensure participants have read these insights, they were required to answer a single-item question within each of these insight panels which we adapted from \citeauthor{samter1987comforting}'s Effectiveness instrument and ~\citeauthor{liu2018should}'s semantic differential~\cite{samter1987comforting, liu2018should}. Fig.~\ref{fig:simulation-interface} shows how this was embedded into the task interface.

\aptLtoX{\begin{figure*}[h!]
        \centering
        \includegraphics[width=\textwidth]{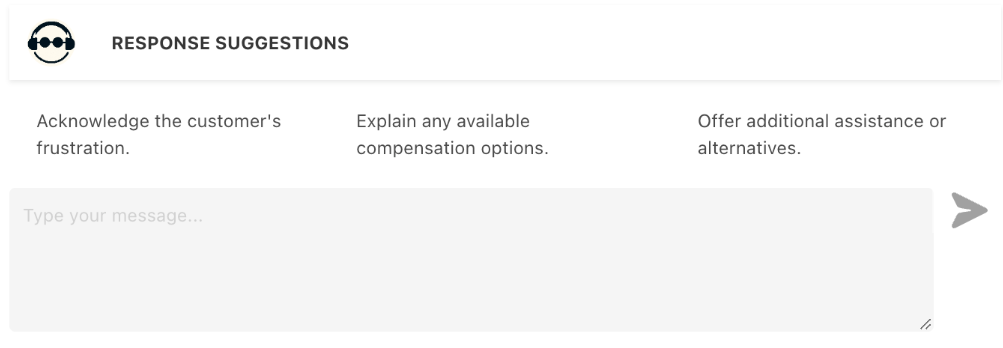}
        \caption{\majorr{Response Field and Response Cues: Aid in continuing conversation}}
        \label{fig:response_field}
        \Description{Screenshot of response field in the simulation exercise.}
    \end{figure*}
    \begin{figure*}[h!]
        \centering
        \includegraphics[width=\textwidth]{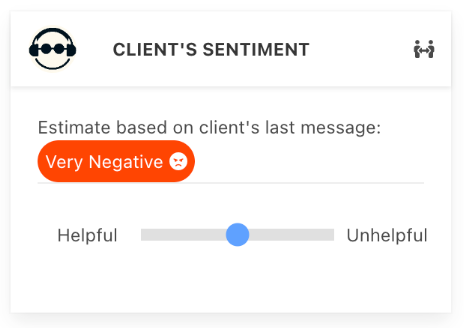}
        \caption{\majorr{\sentiment: Computed using multiple sentiment analysis classifiers}}
        \label{fig:sentiment_pane}
        \Description{Screenshot of the sentiment pane in the simulation exercise.}
\end{figure*}}{\begin{figure*}[h!]
    \begin{minipage}[b]{0.66\textwidth}
        \centering
        \includegraphics[width=\textwidth]{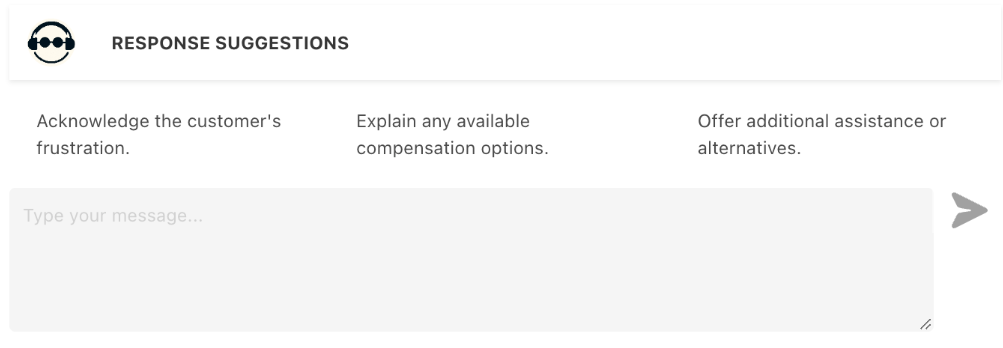}
        \caption{\majorr{Response Field and Response Cues: Aid in continuing conversation}}
        \label{fig:response_field}
        \Description{Screenshot of response field in the simulation exercise.}
    \end{minipage}
    \hfill
    \begin{minipage}[b]{0.32\textwidth}
        \centering
        \includegraphics[width=\textwidth]{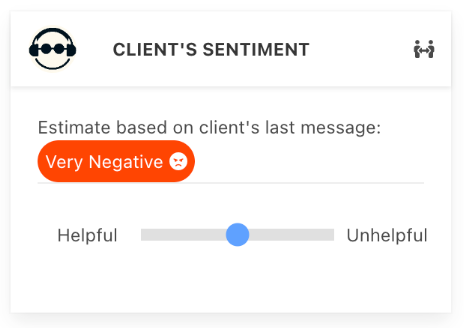}
        \caption{\majorr{\sentiment: Computed using multiple sentiment analysis classifiers}}
        \label{fig:sentiment_pane}
        \Description{Screenshot of the sentiment pane in the simulation exercise.}
    \end{minipage}
\end{figure*}}

\aptLtoX{\begin{figure*}[h!]
        \centering
        \includegraphics[width=\textwidth]{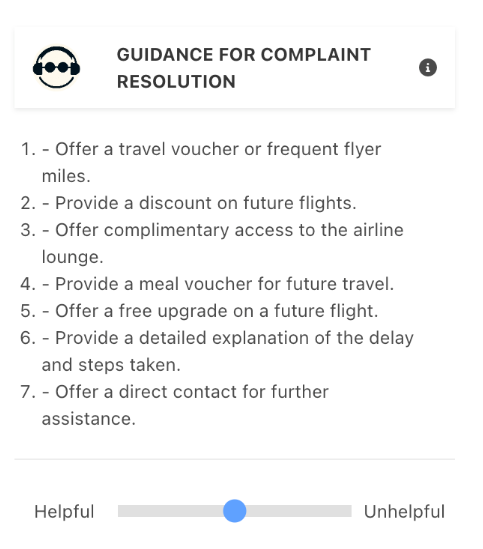}
        \caption{\majorr{\infoguide: Guidelines to resolve the complaint}}
        \label{fig:info-guide_pane}
        \Description{Screenshot of the troubleshooting guidelines pane in the simulation exercise.}
    \end{figure*}
    \begin{figure*}[h!]
        \centering
        \includegraphics[width=\textwidth]{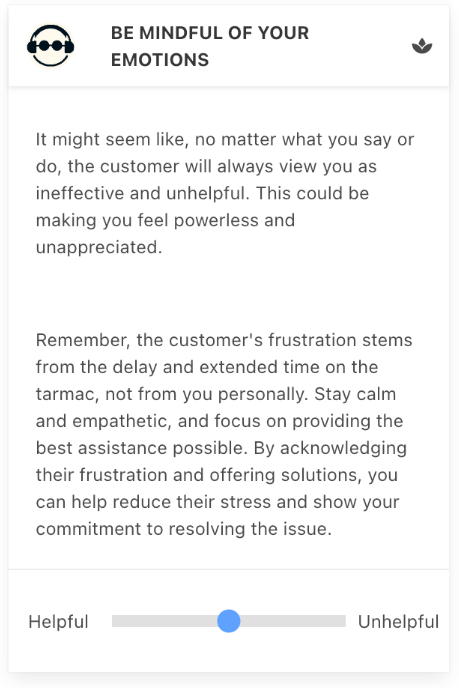}
        \caption{\majorr{\reframe: Generated by \propilot using real data and domain knowledge.}}
        \label{fig:emo-reframe_pane}
        \Description{Screenshot of the reframe pane in the simulation exercise.}
\end{figure*}}{\begin{figure*}[h!]
    \begin{minipage}[b]{0.4\textwidth}
        \centering
        \includegraphics[width=\textwidth]{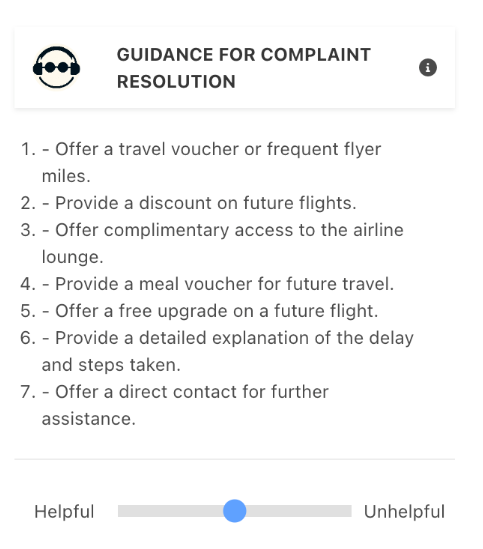}
        \caption{\majorr{\infoguide: Guidelines to resolve the complaint}}
        \label{fig:info-guide_pane}
        \Description{Screenshot of the troubleshooting guidelines pane in the simulation exercise.}
    \end{minipage}
    \hspace{0.05\textwidth}
    \begin{minipage}[b]{0.4\textwidth}
        \centering
        \includegraphics[width=\textwidth]{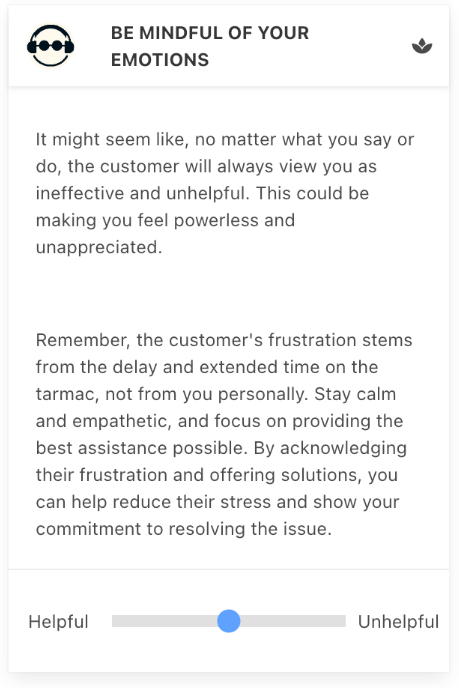}
        \caption{\majorr{\reframe: Generated by \propilot using real data and domain knowledge.}}
        \label{fig:emo-reframe_pane}
        \Description{Screenshot of the reframe pane in the simulation exercise.}
    \end{minipage}
\end{figure*}}

% Following the approach by \cite{liu2018should}, we use a 5-point semantic differential scale.  

\subsubsection{Post-task Measurement and Pre-task Measurement }
\label{si:post-task-measurement}

After participants selected a scenario, participants were required to complete a pre-task survey to establish baseline attitudes. Upon concluding the conversation with each client, the participants were asked to reflect on the entire interaction and complete a post-task survey. The post-task survey contained additional questions, including Q4 (listed below), which was not present in the pre-task survey: 

\textbf{Client Interaction:}
\begin{description}
\small
    \item[Q1.]  To what extent do you agree with the following statements about the client you conversed with (adapted from~\citeauthor{spencer2009angry}\cite{spencer2009angry}):\\ 
    Strongly disagree $\cdot$ Disagree	$\cdot$ Somewhat disagree	$\cdot$ Neither agree nor disagree		$\cdot$ Somewhat agree	$\cdot$ Agree		$\cdot$ Strongly agree
\end{description}
 \begin{enumerate}
\small
     \item The client treated me in a polite manner.   
     \item The client treated me with dignity. 
     \item The client treated me with respect.
 \end{enumerate}

\textbf{Cognitive Demands/Resources.} We adapted two 5-point Likert scale questions from~\citeauthor{demerouti2001job}~\cite{demerouti2001job}:\\
Very low $\cdot$	Low $\cdot$ 		Moderate $\cdot$  	High $\cdot$  	Very High 
 \begin{description}
\small
     \item[Q2a.] In the context of your last conversation, how would you rate the demands on you?  
     \item[Q2b.] In the context of your last conversation, how would you rate the resources available to you? 
 \end{description}

\textbf{Affect.} We adapted two questions from ~\citeauthor{betella2016affective}, scored on a semantic differential scale~\cite{betella2016affective}:

\begin{description}
\small
    \item[Q3.] How do you feel after the conversation? 
    \item [a.] Rate your level of pleasure \\
    Negative-----------|-----------|-----------|-----------|----------- Positive 
    \item [b.] Rate your level of energy \\
    Calm-----------|-----------|-----------|-----------|----------- Excited 
\end{description}

\textbf{Emotional Support.} We adapted the \textit{Effectiveness} and \textit{Supportiveness} scales from ~\citeauthor{liu2018should}~\cite{liu2018should}. Each question was a 5-point semantic differential item; we list the extreme poles below:

\begin{description}
\small
    \item[Q4.]  How did you feel after reading the messages from \propilot? 
\end{description}

\begin{itemize}
\small
    \item Effective / Ineffective 
    \item Helpful / Unhelpful 
    \item Beneficial / Not Beneficial 
    \item Adequate / Inadequate 
    \item Sensitive / Insensitive 
    \item Caring / Uncaring 
    \item Understanding / Not Understanding 
    \item Supportive / Unsupportive 
\end{itemize}

\end{document}